\documentclass[
superscriptaddress,
showpacs,preprintnumbers,nofootinbib,amsmath,amssymb,aps,prd,11pt, groupedaddress]{revtex4-1}
\usepackage{braket}
\usepackage{bm}
\usepackage{dcolumn}
\usepackage{cancel}
\usepackage[dvipdfmx,final]{graphicx}
\input{colordvi.tex}
\usepackage{longtable}
\usepackage{tabularx}
\usepackage{cleveref}
\usepackage{mhchem}
\usepackage{mathtools}
\usepackage[utf8]{inputenc}
\usepackage{color}
\usepackage{amsmath}
\usepackage{graphics}
\usepackage{epstopdf}
\usepackage{graphicx}
\usepackage{subfigure}

\makeatletter
\renewcommand{\fnum@table}{\textbf{\tablename~\thetable}}
\renewcommand{\fnum@figure}{\textbf{\figurename~\thefigure}}
\makeatother

\newcommand {\be}{\begin{equation}}
\newcommand {\ee}{\end{equation}}
\newcommand {\ba}{\begin{eqnarray}}
\newcommand {\ea}{\end{eqnarray}}

\def\epsmutau{\varepsilon_{\mu\tau}}

\begin{document}

\title{Non-Standard Interaction of atmospheric neutrino in future experiments}

\author{Pouya Bakhti  }
\email{pouya\_bakhti@jbnu.ac.kr}
\author{Meshkat Rajaee   }
\email{meshkat@jbnu.ac.kr}
\author{Seodong Shin    }
\email{sshin@jbnu.ac.kr}

\affiliation{Department of Physics, Jeonbuk National University, Jeonju, Jeonbuk 54896, Korea}

\begin{abstract}

We show the prospects of probing neutral-current non-standard interaction (NSI) in the propagation of atmospheric neutrinos in future large-volume neutrino experiments including DUNE, HK, KNO, and ORCA.
For DUNE, we utilize its ability of identifying the tau neutrino event and combine the $\nu_\tau$ appearance with the $\nu_\mu$ disappearance.
Based on our simulated results, the ten years of data taking of the atmospheric neutrinos can enormously improve the bounds on the NSI parameters $\varepsilon_{\mu \tau}, | \varepsilon_{\mu \mu} - \varepsilon_{\tau \tau} |$, $\varepsilon_{e \mu }$, $\varepsilon_{e \tau}$ and $| \varepsilon_{\mu \mu} - \varepsilon_{e e} |$ by a couple of orders of magnitudes.
In addition, we show the expected correlations between the CP-violation phase $\delta_{CP}$ and the NSI parameters $\varepsilon_{e\mu}, \varepsilon_{e\tau}$, and $|\varepsilon_{ee} - \varepsilon_{\mu \mu}|$ and confirm the potentials of DUNE, HK, KNO (combined with HK) in excluding the ``No CP violation" hypothesis at 1$\sigma$, 2$\sigma$, and 3$\sigma$, respectively.
\end{abstract}

\maketitle

\section{Introduction}
\label{sec:intro}

Data from atmospheric neutrinos produced by cosmic-ray interactions in Earth’s atmosphere confirmed neutrino flavor oscillation in 1998 at the Super-Kamiokande (SK) experiment \cite{Super-Kamiokande:1998kpq} and eventually led to the discovery of neutrino oscillations.
The existence of the neutrino oscillations  normally implies the non-zero values of neutrino masses which are non-degenerate among the flavors.
Since it is not expected in the context of the Standard Model (SM), the neutrino oscillation provides one of the most important empirical evidence of new physics beyond the SM (BSM).

In the framework of the standard three-neutrino mixing paradigm, data from various neutrino experiments can be explained with good accuracy. 
The standard paradigm assumes that there are three flavor eigenstates, i.e., $\nu_e$, $\nu_\mu$, and  $\nu_\tau$, which are the weak gauge eigenstates and they are actually superpositions of three mass eigenstates: $\nu_1$, $\nu_2$, and $\nu_3$ with the masses $m_1$, $m_2$, and $m_3$, respectively.
For Dirac neutrinos, the $3 \times 3$ mixing matrix is parametrized by three angles $(\theta_{12},\theta_{13} ,\theta_{23})$ and one CP-violating phase $\delta_{CP}$, while two more phases exist for Majorana neutrinos.
Neutrino flavor oscillation probabilities are the functions of these four phases and the two independent squared mass differences: $\Delta m^2_{21} \equiv m_2^2 - m_1^2$ and $\Delta m^2_{31} \equiv m_3^2 - m_1^2$.
After decades of investigations on these neutrino mixing parameters in various neutrino experiments, we could successfully find the values of the mixing angles, $\Delta m_{21}^2$, and $|\Delta m_{31}^2|$.
Despite this remarkable success, the values of the Dirac CP-violating phase, the neutrino mass ordering (the sign of $\Delta m_{31}^2$), the octant of $\theta_{23}$, and the type of neutrino mass (whether Dirac or Majorana) are still undetermined yet and remained as the main tasks of the upcoming generation of neutrino experiments such as Deep Underground Neutrino Experiment (DUNE) \cite{ DUNE:2020ypp}, Hyper-Kamiokande (HK) \cite{Hyper-Kamiokande:2018ofw}, Korean Neutrino Observatory (KNO) \cite{kno}, Oscillation Research with Cosmics in the Abyss (ORCA) hosted by the Kilometer Cube Neutrino Telescope (KM3NeT) in the Mediterranean Sea \cite{KM3Net:2016zxf} as well as various neutrinoless double beta decay experiments \cite{GERDA:2013vls}.

Although the neutrino oscillation itself is evidence of BSM, the existence of new physics can conversely modify the standard three-neutrino mixing paradigm, which leads to reinterpretations of the oscillation data.
Examples of such extra effects include the non-standard interactions (NSI), the unitarity violation of the mixing, and the existence of light sterile neutrinos.
In this paper, we analyze how much the {\it atmospheric neutrino data in future experiments} can probe the NSI for the first time. 
Note that the NSI operators are the effective general Fermi interaction operators (including the off-diagonal terms) originating from the existence of heavy new physics particles involved in the gauge interactions of the three active neutrinos ($\nu_e$, $\nu_\mu$, and $\nu_\tau$).
The effects of NSI in the other types of 
neutrino oscillation experiments have been studied extensively in the previous literature~\cite{Fornengo:1999zp,
Fornengo:2001pm,Huber:2001zw,Friedland:2005vy,Gonzalez-Garcia:2011vlg,Esmaili:2013fva,Choubey:2014iia,Proceedings:2019qno,Gonzalez-Garcia:1998ryc, Gonzalez-Garcia:2013usa, Coloma:2019mbs,Bakhti:2014pva,Bakhti:2016gic,Bakhti:2016prn,Bakhti:2020fde,Bakhti:2020hbz}.

There are two types of NSI: neutral current NSI (NC-NSI) and charged current NSI (CC-NSI) which are the effective operators with the forms of the neutral current and the charge current weak interactions, respectively.
The NC-NSI affects the oscillation of an active neutrino while propagating through matter and the CC-NSI affects the production and detection of active neutrinos.
As relatively strong model-independent constraints already exist on the CC-NSI compared to the NC-NSI \cite{Biggio:2009nt, Esteban:2019lfo}, we focus on the impacts of the NC-NSI while atmospheric neutrinos are propagating through the Earth matter in this paper.

As our reference future experiments, we consider DUNE, HK, KNO, and ORCA with the ten years of data taking because of the following reasons.
The DUNE far detector has the ability to identify the $\nu_\tau$ signal event by event, in contrast to the other reference experiments, and hence we can combine the $\nu_\tau$ appearance with the $\nu_\mu$ disappearance data.
The HK water Cherenkov light detector is expected to have a fiducial volume about 6 times larger than the DUNE far detector with low enough threshold energy of detecting $\nu_\mu$ events.
The multi-purpose large volume neutrino telescope KNO can provide less background contaminated data due to its 1000 m scale granite overburden. 
The expected design of the KNO detector is quite similar to the HK water Cherenkov light detector, due to its other task as the second far detector of the long-baseline experiment, Tokai-to-Hyper-Kamiokande-to-Korea (T2HKK), 
so that we can easily combine its data with the HK data.~\footnote{Here, we do not show the effect of the background rejection in KNO and leave it to future work.}
Compared to the other experiments, ORCA has an overwhelming size of fiducial volume about 150 times larger than the DUNE far detector.

In this paper, we consider the neutrino energy range from 1 GeV to 200 GeV since the flux of atmospheric neutrino out of this energy range either drops rapidly or depends highly on the model details~\cite{Honda:2006qj}.
From the neutrinos in the high energy range of $15 - 200$ GeV, we obtain new prospects in probing the NSI parameters $\varepsilon_{\mu \tau}$ and $|\varepsilon_{\mu \mu} - \varepsilon_{\tau \tau}|$.
From the neutrinos in the low energy range of $1-15$ GeV, on the other hand, we probe $\delta_{CP}$, $\varepsilon_{e\mu}$, $\varepsilon_{e\tau}$, and $|\varepsilon_{ee} - \varepsilon_{\mu \mu}|$.

This paper is organized as follows.
The detailed formalism of the NSI framework and our analysis strategies according to the expectations of the oscillation probabilities are introduced in Sec.~\ref{sec:anal}.
The future sensitivities of the reference experiments in probing the NSI parameters are shown in the categories of the high energy range (15 - 200 GeV) and the low energy range (1 - 15 GeV) separately in Sec.~\ref{sec:results}.
We summarize the analysis results and discuss future prospects in Sec.~\ref{sec:conclusions}.

\section{Analysis strategies}
\label{sec:anal}

\subsection{Formalism}

In this section, we review the formalism of the NSI and introduce the conventions used in this analysis.
Neutral-current NSI  at low energies can be described via the effective four-fermion dimension-six operators~\cite{Wolfenstein:1977ue}:
\begin{align}
{\mathcal L}_{\rm NC-NSI} = -2\sqrt 2 \, G_F \,\varepsilon^{f}_{\alpha\beta, C}
\, (\bar\nu_{\alpha}\gamma^\rho P_L\nu_{\beta}) \,
(\bar f \gamma_\rho P_C f) \, ,
\end{align}
where $G_{F}$ is the Fermi constant and $\varepsilon^{f}_{\alpha \beta}$ describes the strength of NSI for each matter fermion $f \in \{e,u,d\}$ with $\alpha, \beta \in \{e,\mu,\tau \}$ being the neutrino flavor. 
The chirality projection matrix $P_C = (1 \mp \gamma^5)/2$ for the left and right-handed gauge interactions, respectively, is included.
Neutrino propagation through matter is controlled by the NSI parameter:
\begin{align}
\varepsilon_{\alpha\beta} \equiv \sum_{f=e,u,d} \left(\varepsilon_{\alpha\beta}^{fL}
+ \varepsilon_{\alpha\beta}^{fR}\right) \,\,\frac{N_f}{n_e}
\equiv \sum_{f=e,u,d} \varepsilon_{\alpha\beta}^{f}\,\, \frac{N_f}{N_e}\,,
\end{align}
where $N_f$ is the number density of fermion $f$. 
Assuming $N_u \approx N_d \approx 3 N_e$, the above parameter becomes
\begin{align}
    \varepsilon_{\alpha\beta} \approx \varepsilon^e_{\alpha\beta}
+ 3 \, \varepsilon^u_{\alpha\beta} + 3 \, \varepsilon^d_{\alpha\beta}\,.
\end{align}

The effective Hamiltonian of neutrino
propagation through matter in the presence of NSI is 
\begin{align}
H_{\rm eff} = \frac{1}{2E} \, U \left(\begin{array}{ccc}
0 & 0 & 0 \\
0 & \Delta m^2_{21} & 0 \\
0 & 0 & \Delta m^2_{31}\end{array}
\right) U^\dag + V_{\rm CC}\left(\begin{array}{ccc}
1 + \varepsilon_{ee} & \varepsilon_{e\mu} & \varepsilon_{e\tau} \\
\varepsilon_{e\mu}^* & \varepsilon_{\mu\mu} & \varepsilon_{\mu\tau} \\
\varepsilon_{e\tau}^* & \varepsilon_{\mu\tau}^* & \varepsilon_{\tau\tau}
\end{array}\right)\,,\label{h}
\end{align}
where $U$ is the Pontecorvo-Maki-Nakagawa-Sakata (PMNS) matrix \cite{Pontecorvo:1957qd,Maki:1962lba}, $\Delta m^2_{ij} \equiv m_i^2 - m_j^2$ are the mass-squared differences, and $V_{CC} \equiv \pm \sqrt{2} G_F N_e$ is the effective matter
potential describing coherent elastic forward scattering of neutrinos in ordinary matter,
where $\pm$ denotes the cases for neutrinos and anti-neutrinos, respectively.
The oscillation probabilities in the presence of NSI are obtained by numerically solving Eq.~(\ref{h}).
Note that the oscillations are insensitive to an overall shift to the eigenvalues of $H_{\rm eff}$ in Eq.~(\ref{h}), i.e., subtraction by $\varepsilon_{\mu \mu} I_{3 \times 3}$ conventionally where $I_{3 \times 3}$ is the $3\times 3$ identity matrix~\cite{DeGouvea:2019kea}.
This renders us to represent the oscillations in terms of $|\varepsilon_{ee} - \varepsilon_{\mu \mu}|$ and $|\varepsilon_{\mu \mu} - \varepsilon_{\tau \tau}|$.

\subsection{Oscillation probabilities}

\begin{figure}[h]
\begin{center}
\includegraphics[width=0.327 \textwidth]{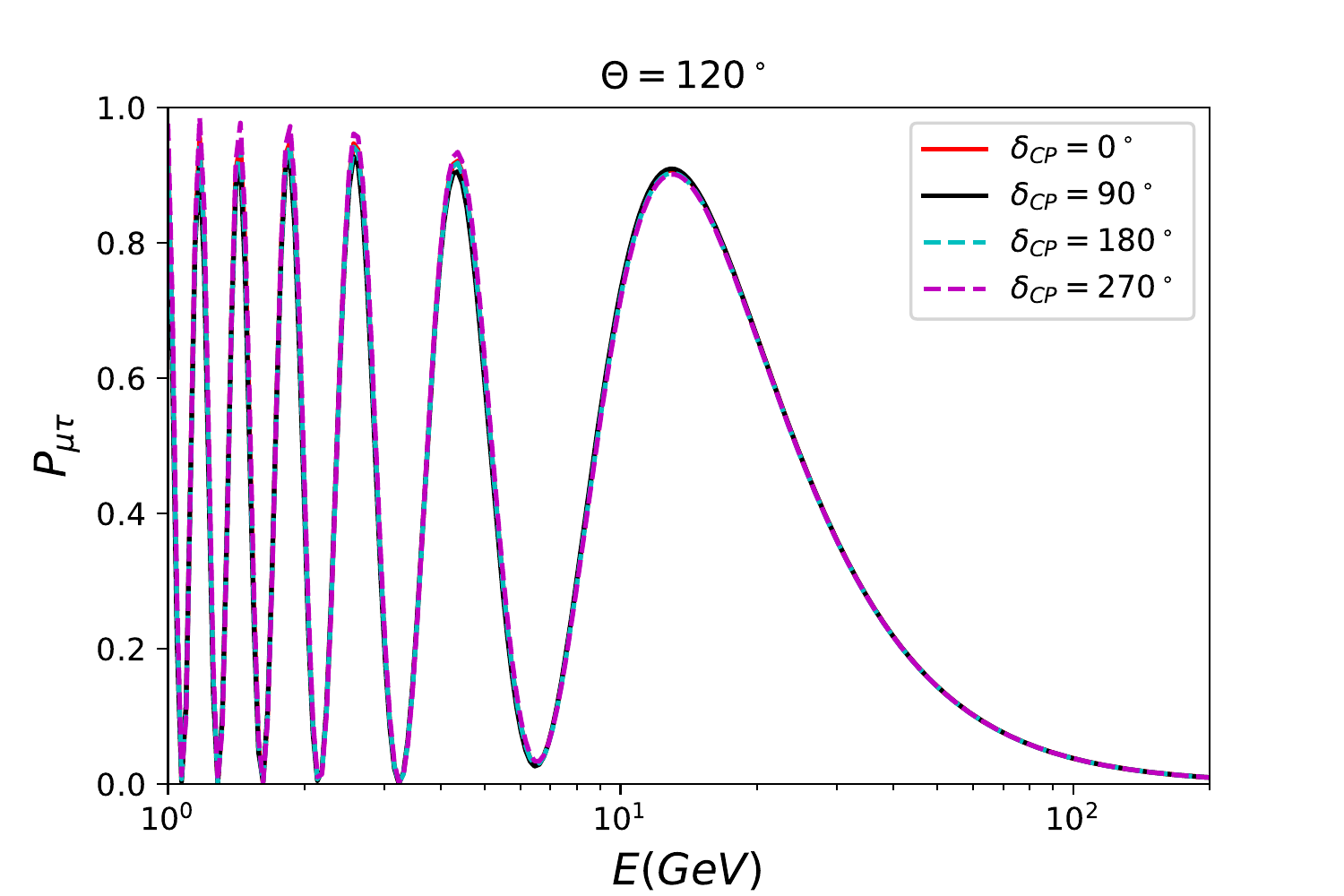}
\includegraphics[width=0.327 \textwidth]{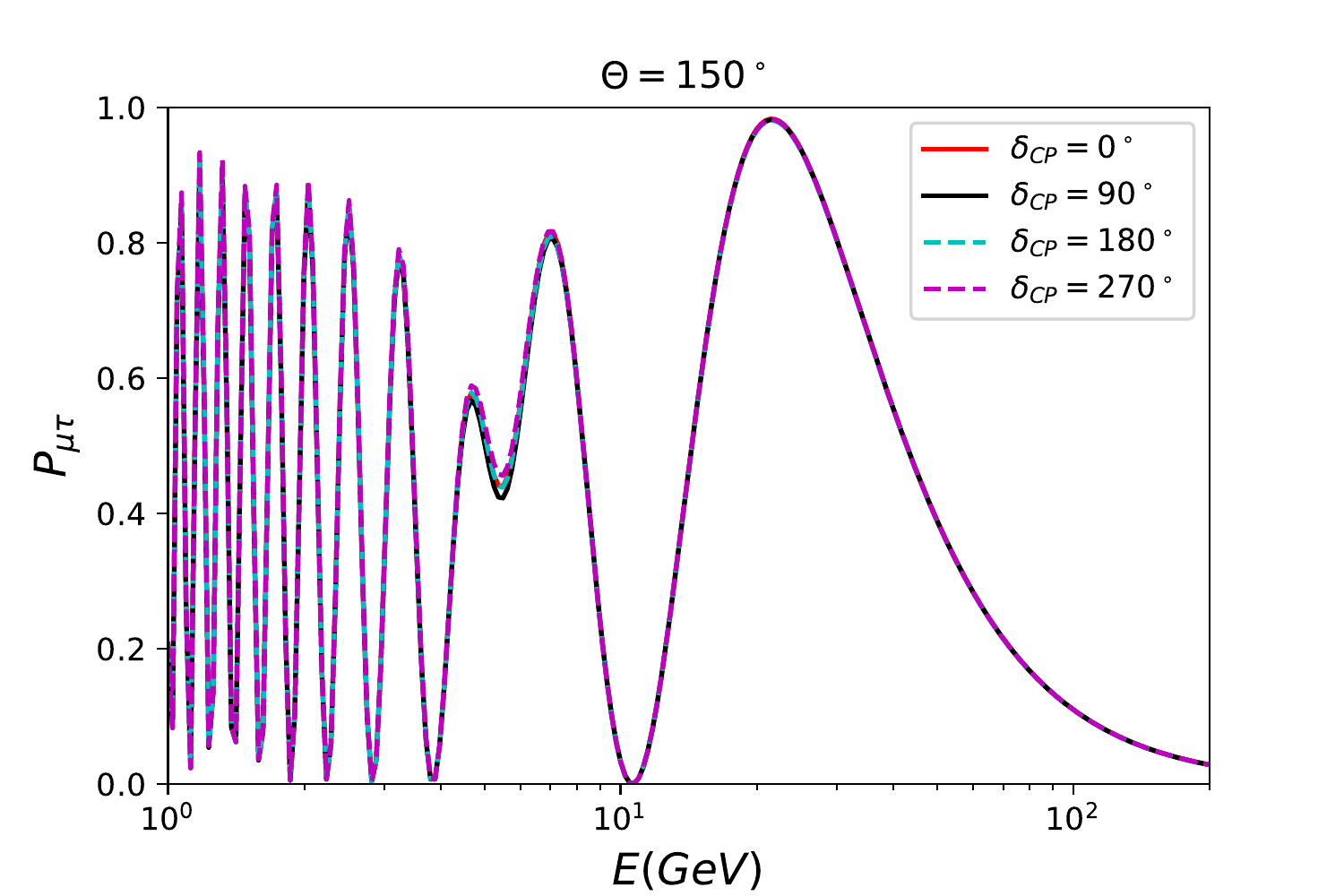}
\includegraphics[width=0.327 \textwidth]{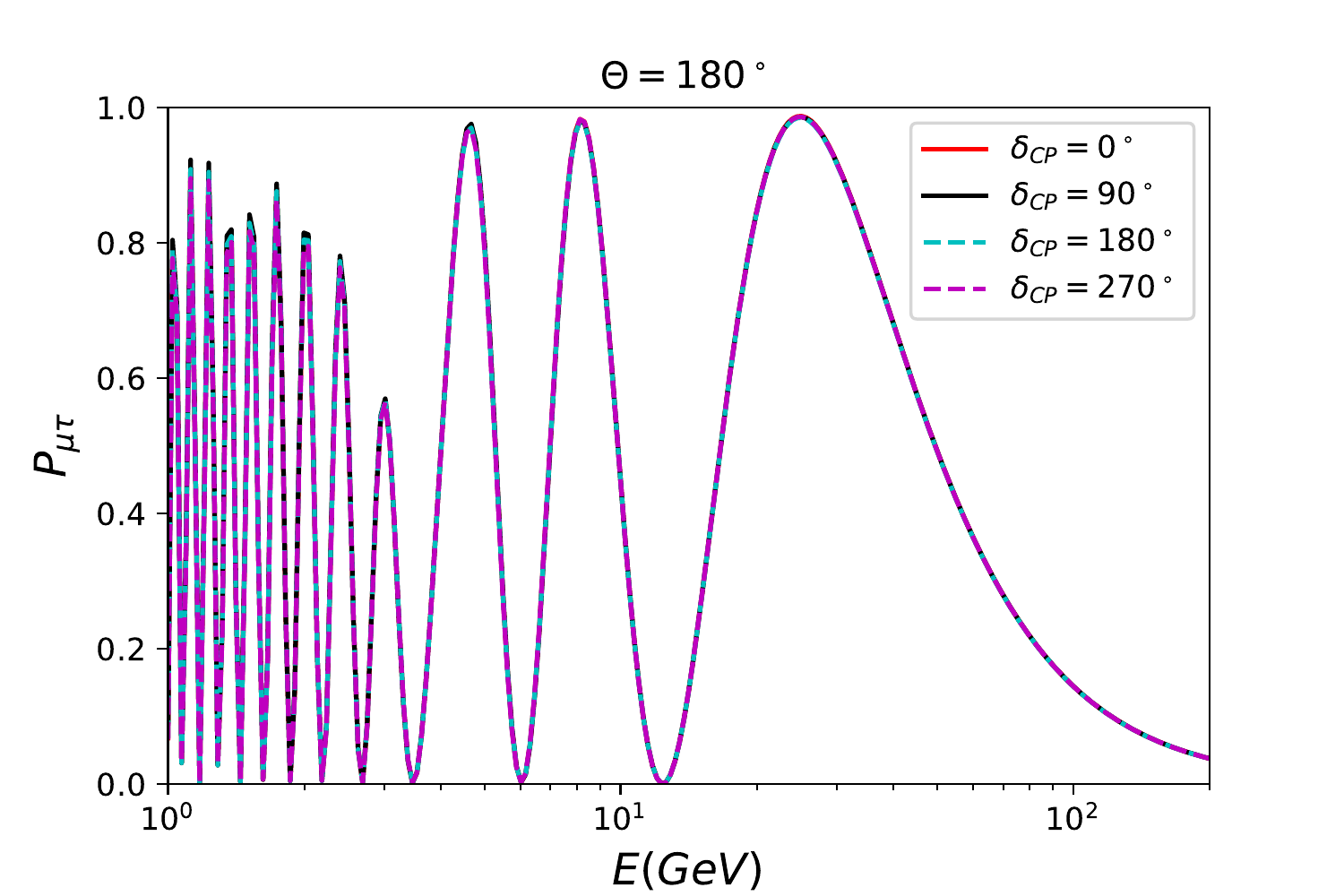}
\end{center}
\caption{\label{deltaa} $P_{\mu\tau}$ as a function of energy for different values of $\delta_{CP}$ for three different fixed values of the zenith angle: $\Theta = 120^\circ$, $150^\circ$ and $180^\circ$  for $\Delta m_{31}^2 = 2.51\times10^{-3}$ and $\theta_{23} = 49^\circ$, a best-fitted value including the Super-Kamiokande atmospheric data.~\cite{Esteban:2020cvm}.
}
\end{figure}

Strong experimental constraints from the meson and muon decays apply to the charged current NSI operators~\cite{Davidson:2003ha,Biggio:2009nt}, while the neutral current NSI operators are rather weakly constrained as summarized in Ref.~\cite{Esteban:2020itz}.
A global analysis of oscillation data on $|\epsmutau|$ provides the constraint: $|\varepsilon_{\mu\tau}|\lesssim 0.02$ at 2$\sigma$ C.L., considering contributions to NSI from only up and down quarks~\cite{Esteban:2018ppq}.
Note that the individual bounds from the large volume neutrino experiments can be stronger: $-0.006 \lesssim \varepsilon_{\mu\tau} \lesssim 0.0054$ from IceCube~\cite{Salvado:2018edx},  $-0.0067 \lesssim \varepsilon_{\mu\tau} \lesssim 0.0081$ from DeepCore~\cite{IceCube:2017zcu}, $|\varepsilon_{\mu\tau}| \lesssim 0.011$ from Super-Kamiokande (SK), all of which are at at $90 \%$ C.L..
On the other hand, the constraints on $|\varepsilon_{\mu\mu}-\varepsilon_{\tau \tau}|$ are less stringent.
The global analysis in Ref.~\cite{Esteban:2018ppq} expects $|\varepsilon_{\mu\mu}-\varepsilon_{\tau \tau}| \lesssim 0.05$ or weaker.

Interestingly, it is easy to constrain $\varepsilon_{\mu \tau}$ and
$| \varepsilon_{\mu \mu} - \varepsilon_{\tau \tau} |$ by considering the $\nu_\mu$ disappearance and the $\nu_\tau$ appearance for $E > 15$ GeV.
This is because the oscillation probabilities can be obtained in the $2\nu$ approximated system in the energy range well above the $\theta_{13}$ resonance which occurs around $2-10$ GeV depending on the averaged density in the Earth core or mantle~\cite{Esmaili:2013fva}; the relevant NSI parameters are just $\varepsilon_{\mu \tau}$ and $| \varepsilon_{\mu \mu} - \varepsilon_{\tau \tau} |$, and $P_{e e } \sim 1$.
We justify this argument by showing the effect of these NSI parameters on the oscillation probability of $\nu_\mu \to \nu_\tau$, defined as $P_{\mu \tau}$, from full numerical calculations by solving the three-generation differential equation of motion for the neutrinos coming from the atmosphere.
In our calculations, the Preliminary Reference Earth Model (PREM) profile~\cite{Dziewonski:1981xy} taking into account the averaged density is used.
We set the related NSI parameters to be non-zero, one at a time, and the NSI phases to zero for simplicity.

Figure~\ref{deltaa} shows the oscillation probability $P_{\mu\tau}$ as a function of energy for different values of $\delta_{CP}$ with the atmospheric $\nu_\mu$ having the zenith angle $\Theta = 120^\circ$, $\Theta = 150^\circ$ and $\Theta = 180^\circ$, from the left to the right panels, respectively.
We fix the other neutrino oscillation parameters as given in Ref. \cite{Esteban:2020cvm} including the SK atmospheric data.

It can be observed that the oscillation probability is not sensitive to the value of $\delta_{CP}$ for energies more than 15 GeV. 
This result is easily expected because the three flavor neutrino oscillation probability can be approximated by the  two flavor form; there is no CP violation with only two flavors. 
We show the energy range $E \le 200$ GeV since the flux of the atmospheric neutrino reduces as $E^{-3.7}$~\cite{Okumura:2018loq}.

\begin{figure}[h]
\begin{center}
\includegraphics[width=0.48\textwidth]{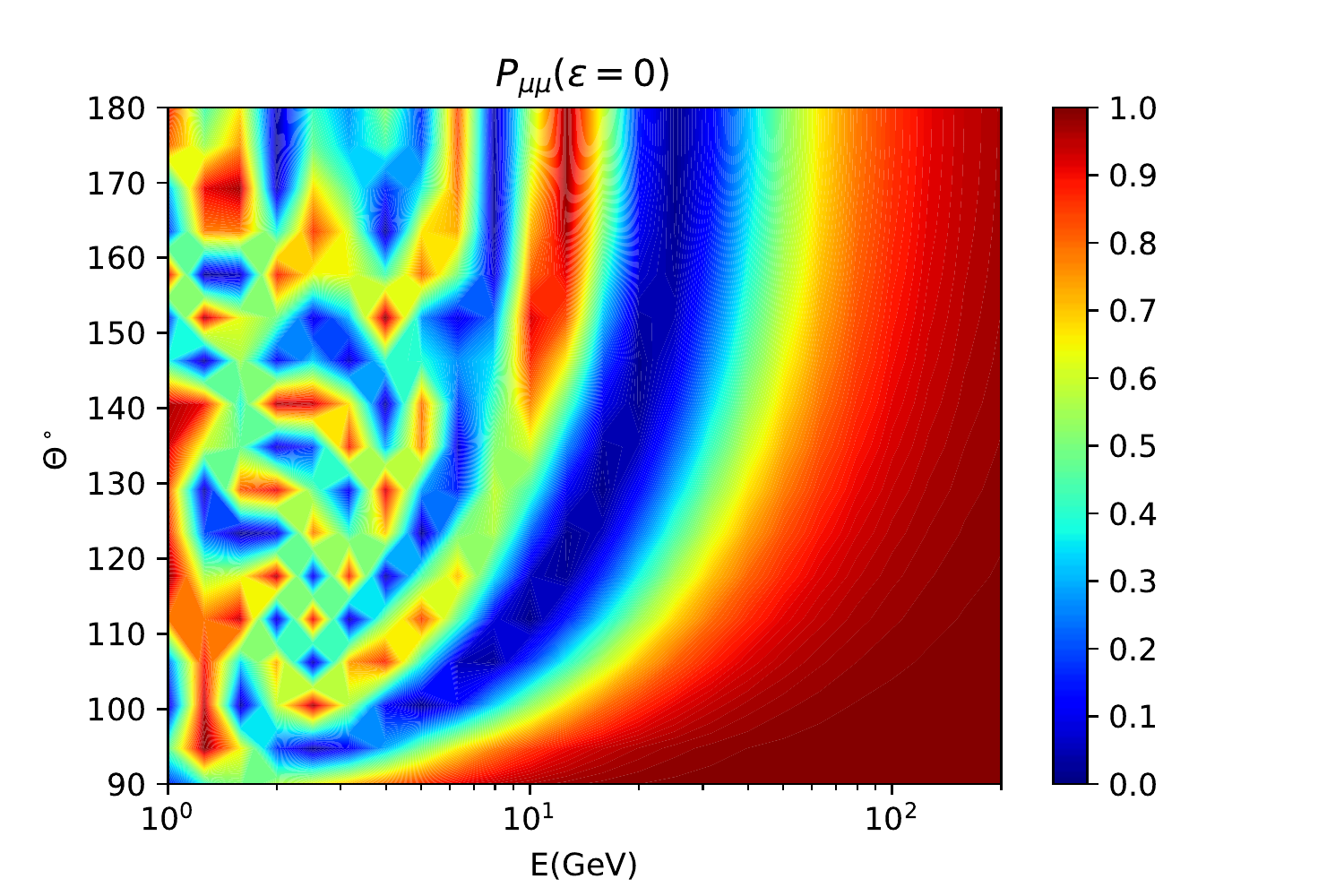}\\
\includegraphics[width=0.45 \textwidth]{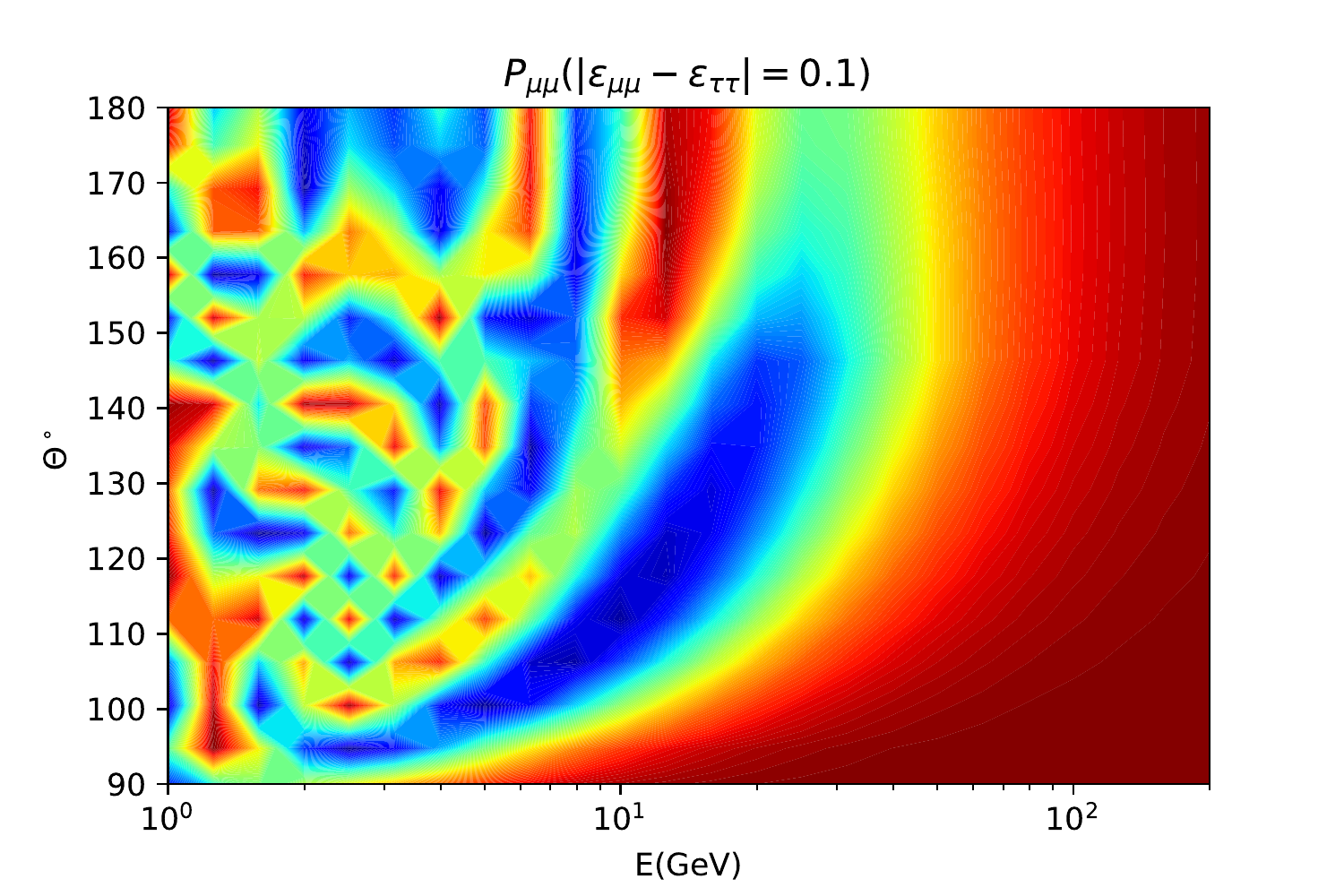}
\includegraphics[width=0.45 \textwidth]{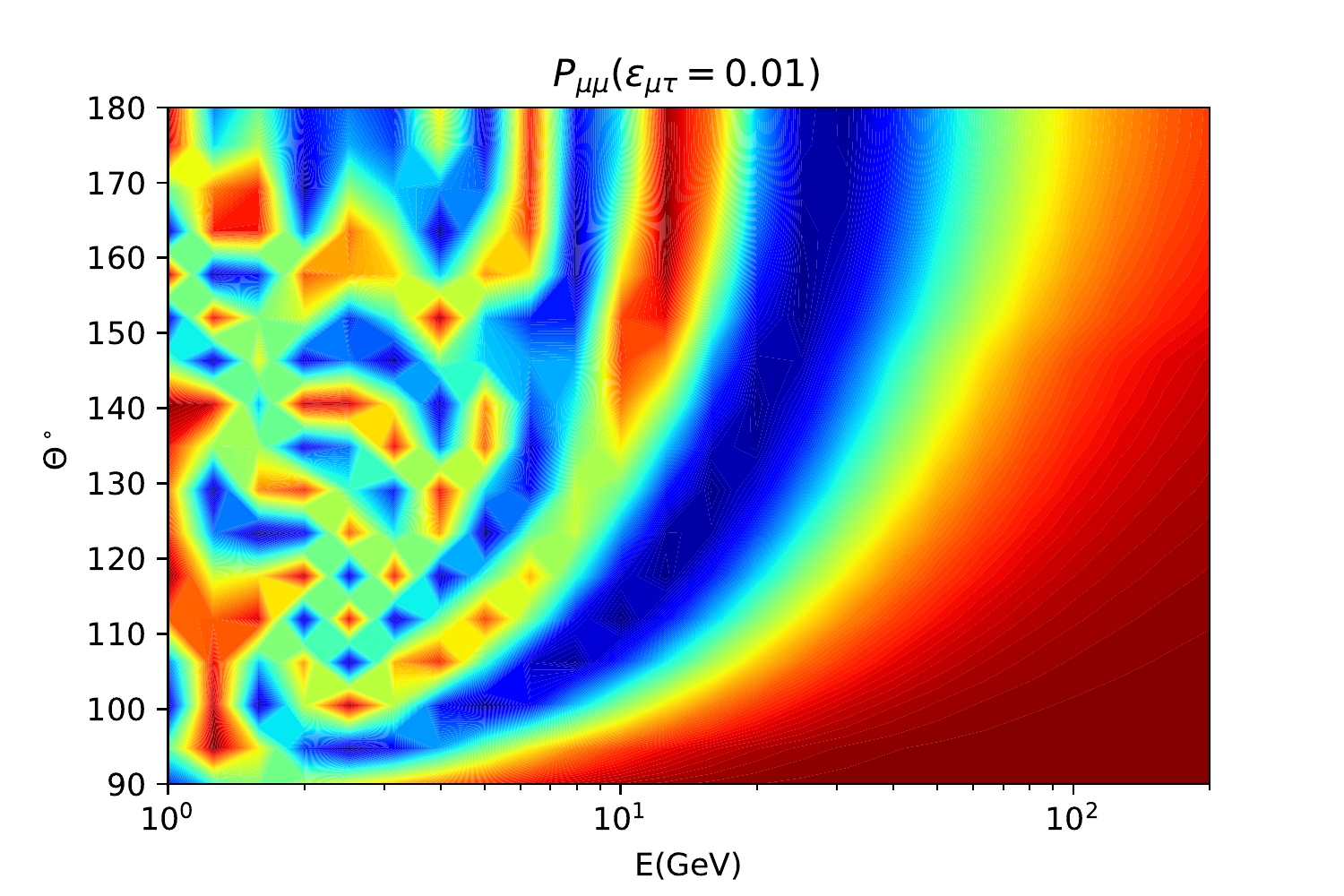}
\end{center}
\caption{\label{oscmm} Oscillograms for the probability $P_{\mu \mu}$ assuming $\delta_{CP}=0^\circ$.
We have considered known normal mass ordering,
and the oscillation parameters are set as given in \cite{Esteban:2020cvm}. The upper panel indicates $P_{\mu \mu}$ in the absence of NSI parameters. In the lower left (right) panel we have assumed non-zero $| \varepsilon_{\mu \mu} - \varepsilon_{\tau \tau} |$ ($\varepsilon_{\mu \tau}$) while setting other NSI parameters to zero.}
\end{figure}

\begin{figure}
\begin{center}
\includegraphics[width=0.48\textwidth]{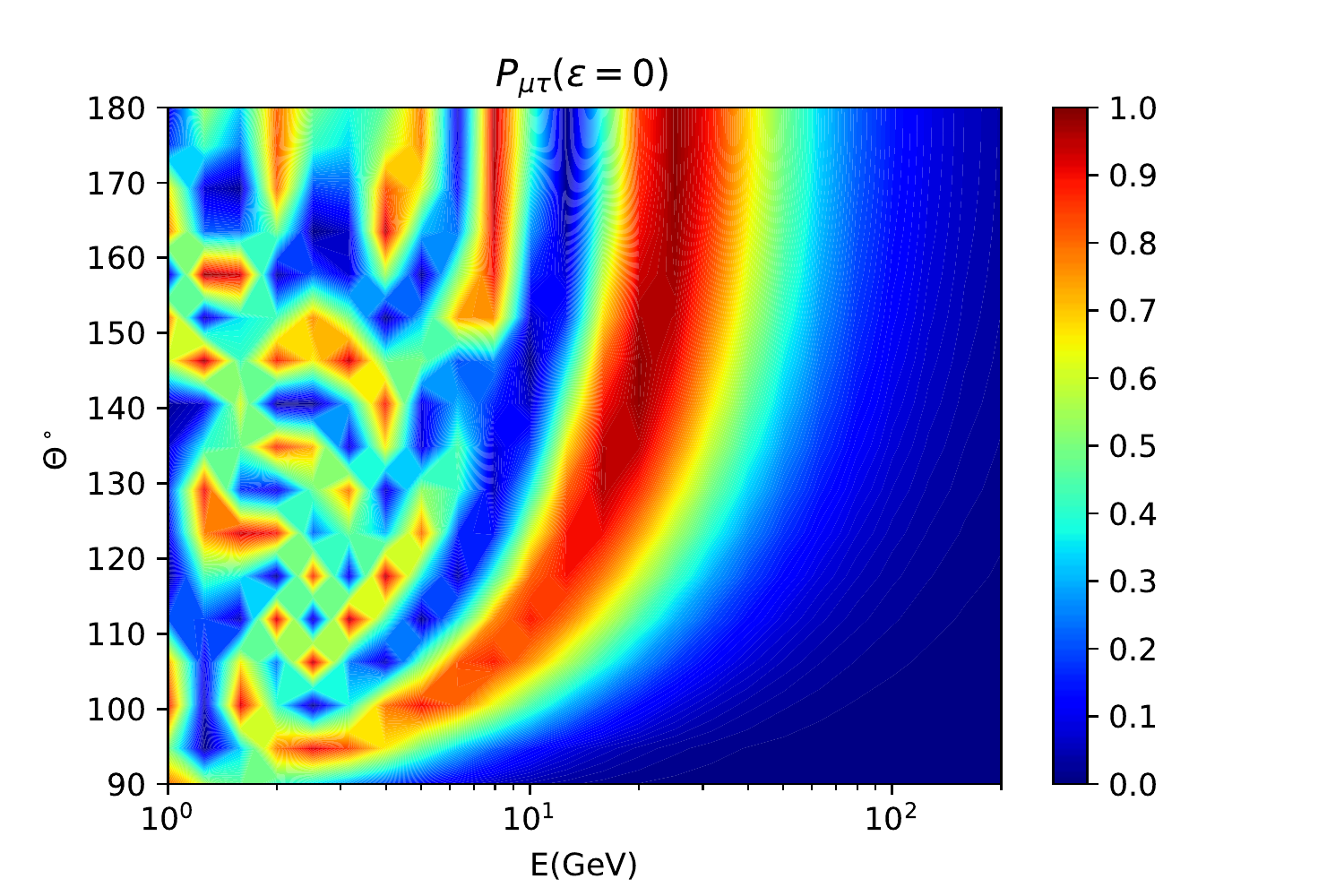}\\
\includegraphics[width=0.45 \textwidth]{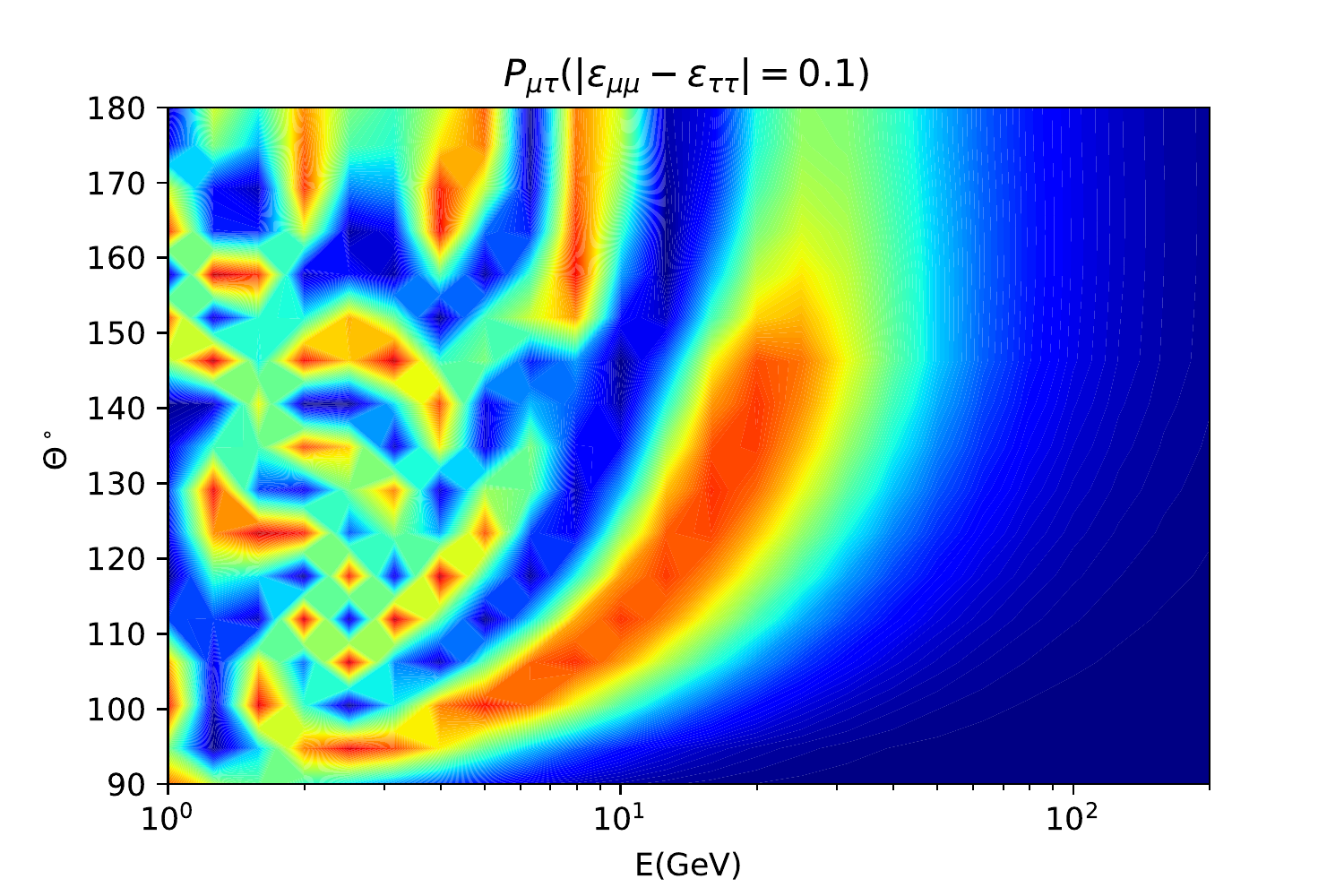}
\includegraphics[width=0.45 \textwidth]{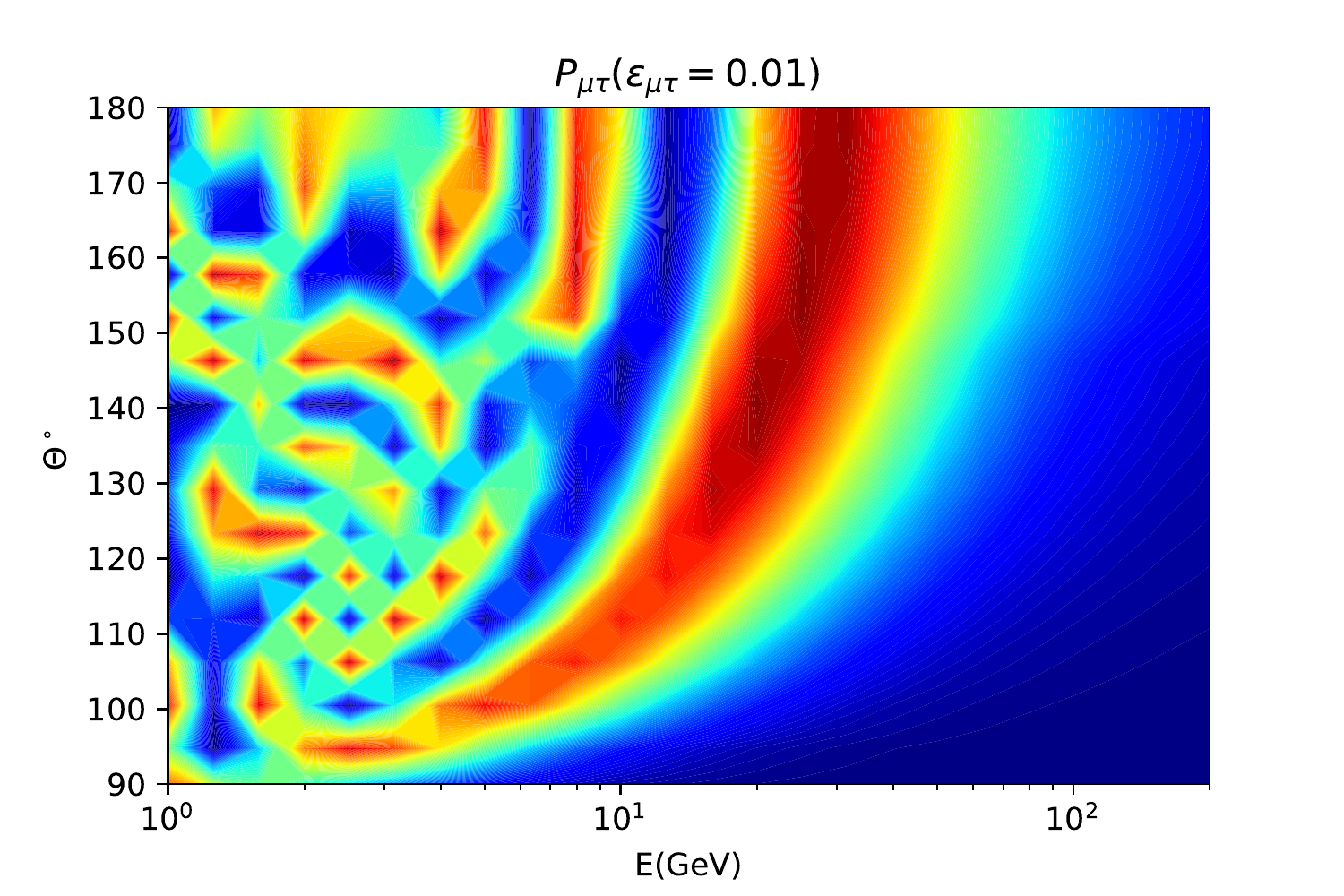}
\end{center}
\caption{\label{oscmt} Oscillograms for the probability $P_{\mu \tau}$ assuming $\delta_{CP}=0^\circ$.
We have considered known normal mass ordering,
and the oscillation parameters are set as given in \cite{Esteban:2020cvm}. In the lower left (right) panel we have assumed non-zero $| \varepsilon_{\mu \mu} - \varepsilon_{\tau \tau} |$ ($\varepsilon_{\mu \tau}$) while setting other NSI parameters to zero.
}
\end{figure}

We now concretely show the dependence of the probabilities $P_{\mu \mu}$ and $P_{\mu \tau}$ on the NSI parameters $\varepsilon_{\mu \tau}$ and $| \varepsilon_{\mu \mu} - \varepsilon_{\tau \tau} |$ for the zenith angle $\Theta \in [90^\circ, 180^\circ]$.
Figures~\ref{oscmm} and \ref{oscmt} are the oscillograms for the oscillation probabilities $P_{\mu \mu}$ and $P_{\mu \tau}$, respectively, assuming $\delta_{CP}=0^\circ$ and the normal mass ordering.
All the oscillation parameters are taken from Ref.~\cite{Esteban:2020cvm}.
In the upper panels of the figures, we turn off all the NSI parameters.
On the other hand, the lower left (right) panels are obtained for $| \varepsilon_{\mu \mu} - \varepsilon_{\tau \tau } | = 0.1$ ($\varepsilon_{\mu \tau} = 0.01$), while setting the other NSI parameters to zero. 
The probabilities $P_{\mu \mu}$ and $P_{\mu \tau}$ governing the $\nu_\mu$ disappearance and $\nu_\tau$ appearance, respectively, are anti-correlated in the figures, as expected.
Interestingly, one can observe the huge effect of turning on the NSI parameters in the oscillation probabilities in the parameter region of high energy $E \gtrsim 15$ GeV and large zenith angle $\Theta \gtrsim 110^\circ$.
For large zenith angels, the atmospheric neutrinos pass through the deep mantle and the core of the Earth where the density is higher, and in consequence, the matter effect is more significant.

Notice that the change of the oscillation probabilities is more sensitive to $\varepsilon_{\mu \tau}$ than $| \varepsilon_{\mu \mu} - \varepsilon_{\tau \tau} |$.
This is because the crucial oscillation channel $\nu_\mu \to \nu_\tau$,  (likewise for the anti-neutrinos) depends more strongly on $|\varepsilon_{\mu \tau}|$ than $|\varepsilon_{\mu \mu} - \varepsilon_{\tau \tau}|$. 
For the two neutrino oscillation scheme, the two parameters controlling the oscillation probabilities, $\theta_{23}^m$ and $\Delta m_{31,m}^2$, in the presence of those NSI parameters can be expressed as~\cite{Agarwalla:2021zfr}
\begin{align}
    \Delta m_{31,m}^2 &\approx \Delta m_{31}^2 \left(1+2\varepsilon_{\mu\tau}A_{CC}+\frac{1}{2}\frac{|\varepsilon_{\mu\mu}-\varepsilon_{\tau\tau}|^2A_{CC}^2}{1+\varepsilon_{\mu\tau}A_{CC}}\right)\,, \\
    \sin2\theta_{23}^m &=\frac{\sin2\theta_{23}(1+2A_{CC}\varepsilon_{\mu\tau})^2}{(A_{CC}|\varepsilon_{\mu\mu}-\varepsilon_{\tau\tau}|)^2+(1+2\varepsilon_{\mu\tau}A_{CC})^2} 
    \simeq    1 - 16 A_{CC}^2 \varepsilon_{\mu \tau}^2 - A_{CC}^2 |\varepsilon_{\mu \mu} - \varepsilon_{\tau \tau}|^2   \,,
\end{align} \label{theta23}
where $A_{CC} = 2EV_{CC}/\Delta m_{31}^2$ for the neutrino energy $E$ and $\theta_{23}$ is a mixing angle in the standard parametrization of the PMNS matrix.
From the above expression for $\Delta m_{31,m}^2$, we can observe that $\varepsilon_{\mu \tau}$ appears in first order while $|\varepsilon_{\mu \mu} - \varepsilon_{\tau \tau}|$ appears in the second order. Moreover, for  effective angle in matter ($\sin2\theta_{23}^m$),  the dependence on $\varepsilon_{\mu \tau}^2$ is 16 times larger than $|\varepsilon_{\mu \mu} - \varepsilon_{\tau \tau}|^2$.

\begin{figure}
\begin{center}
\includegraphics[width=0.44\textwidth]{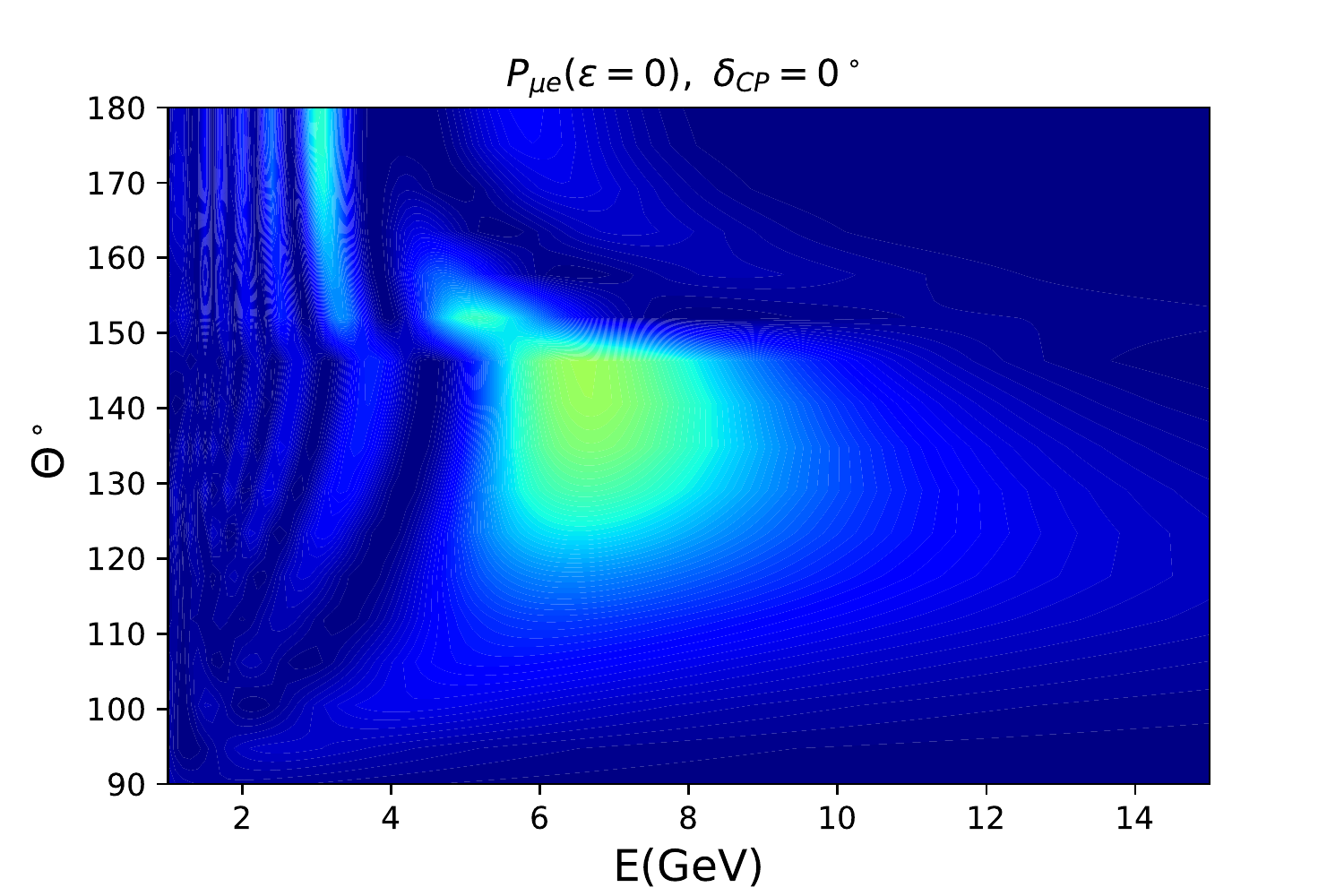}
\includegraphics[width=0.44\textwidth]{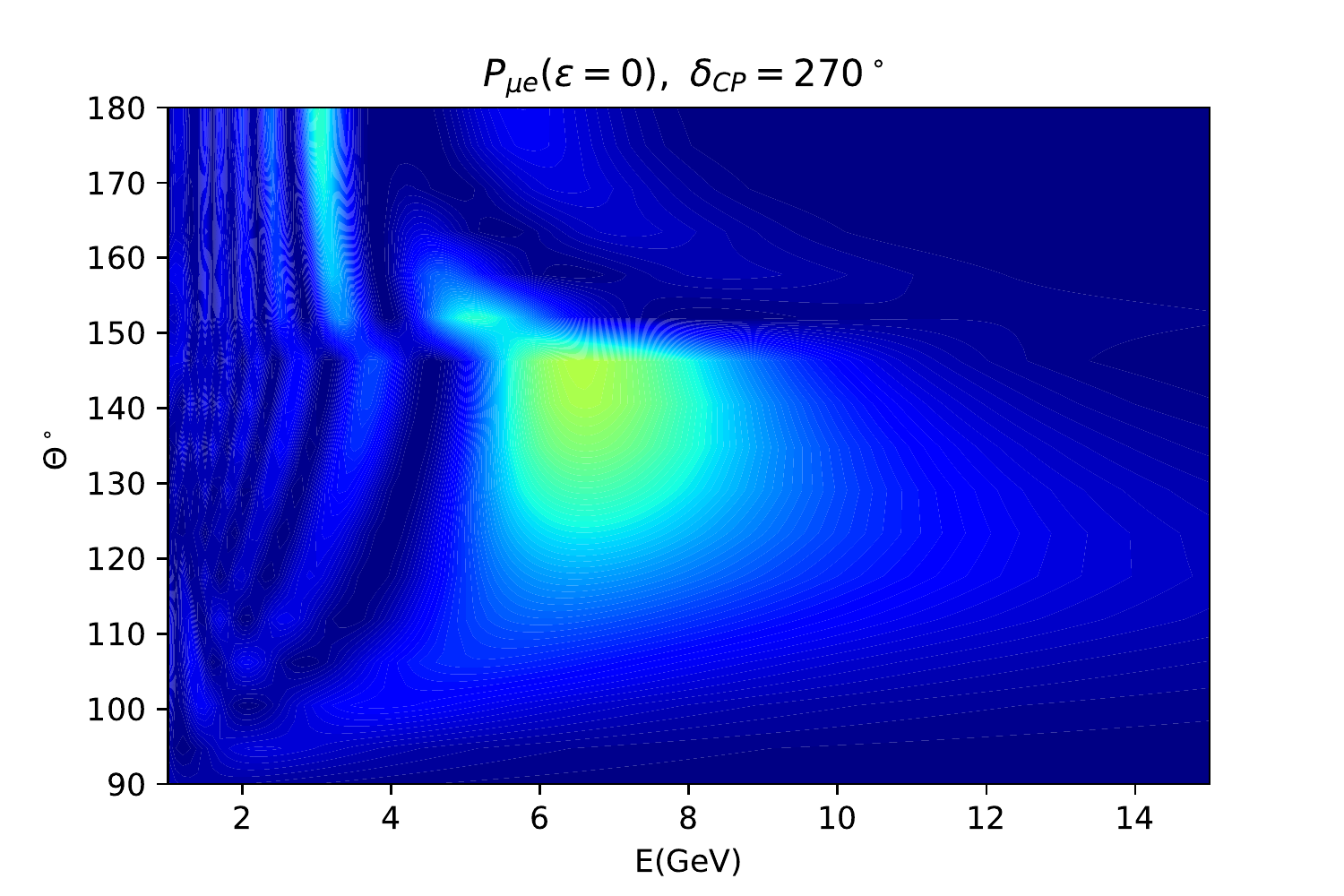}
\includegraphics[width=0.44\textwidth]{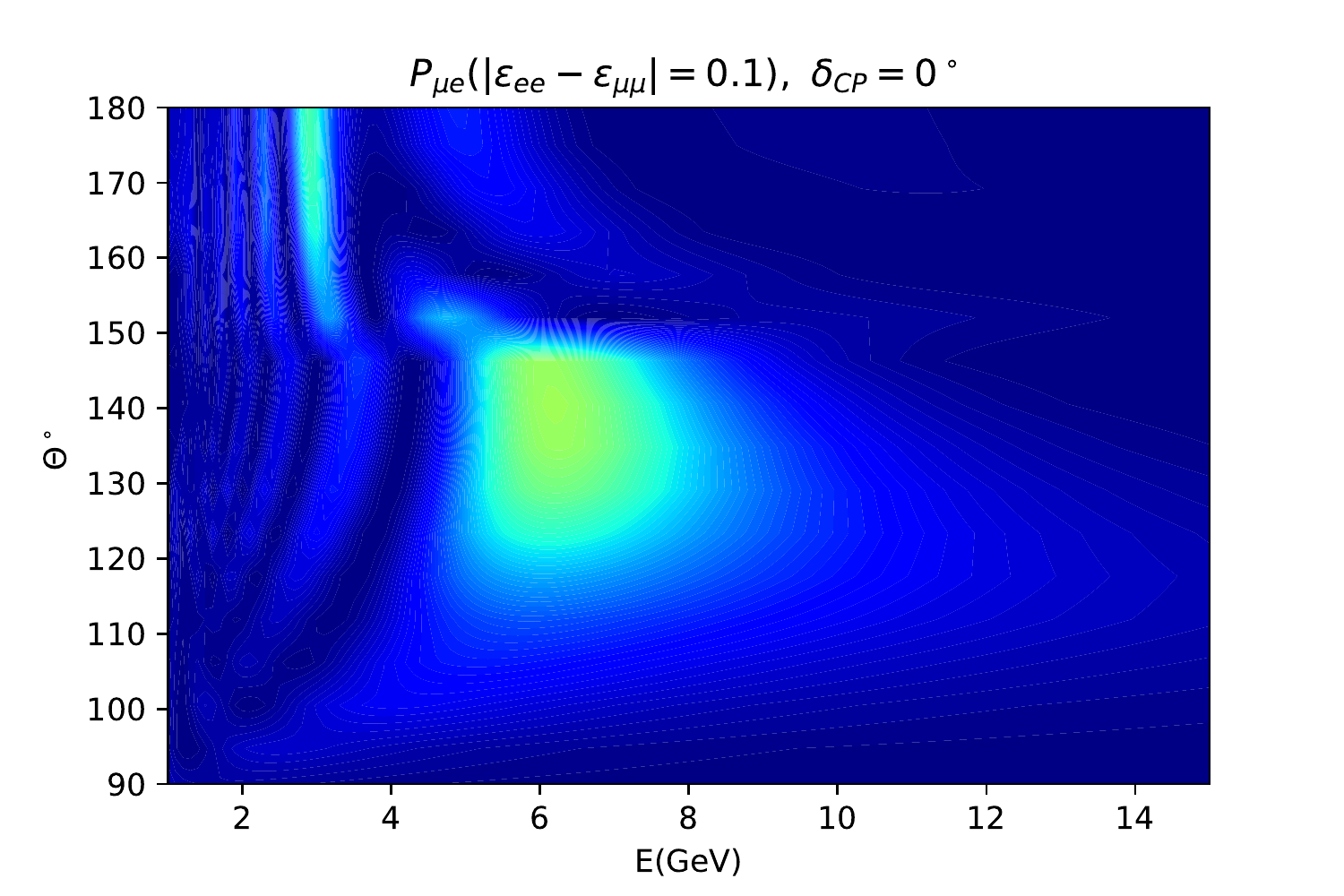}
\includegraphics[width=0.44\textwidth]{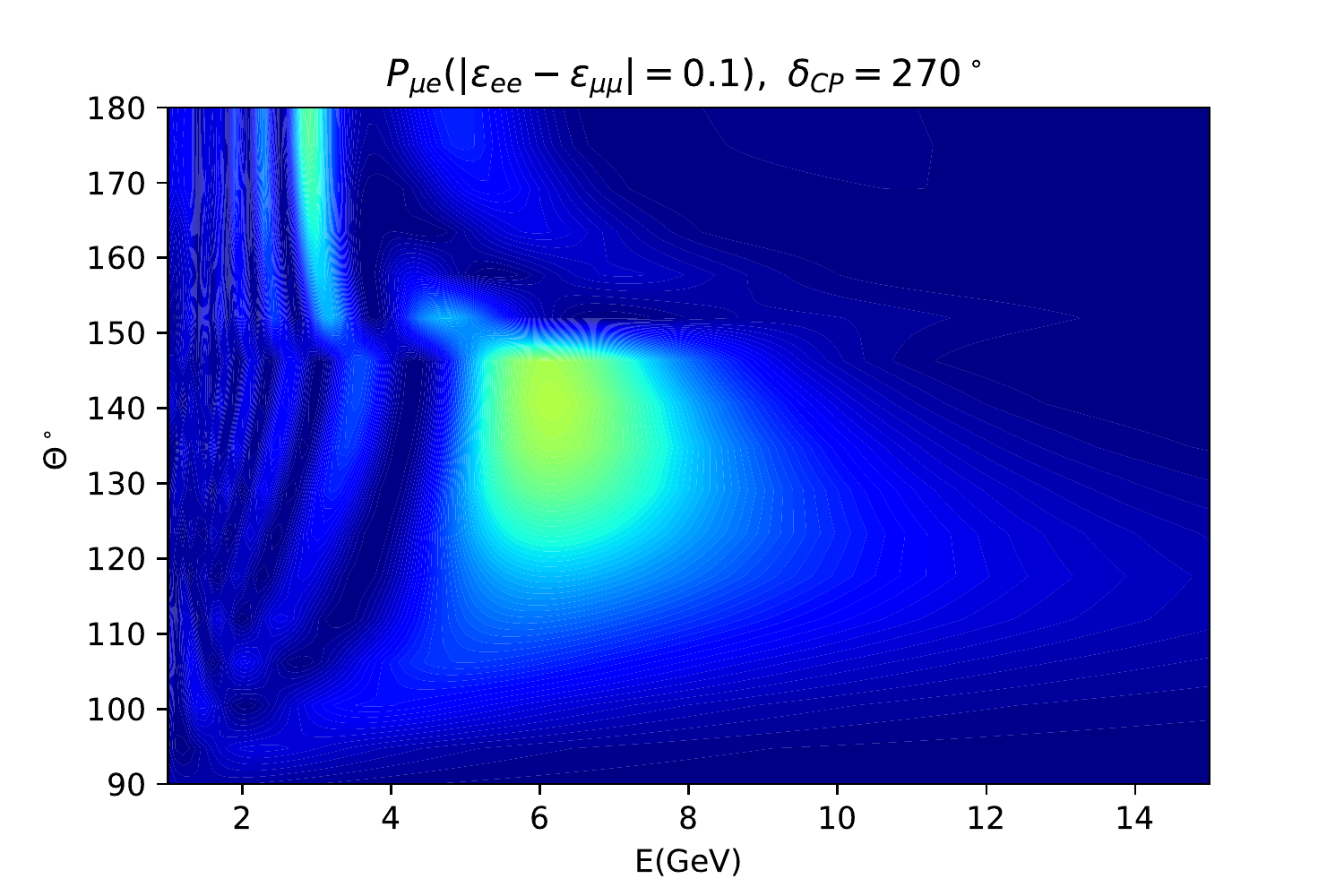}
\includegraphics[width=0.44\textwidth]{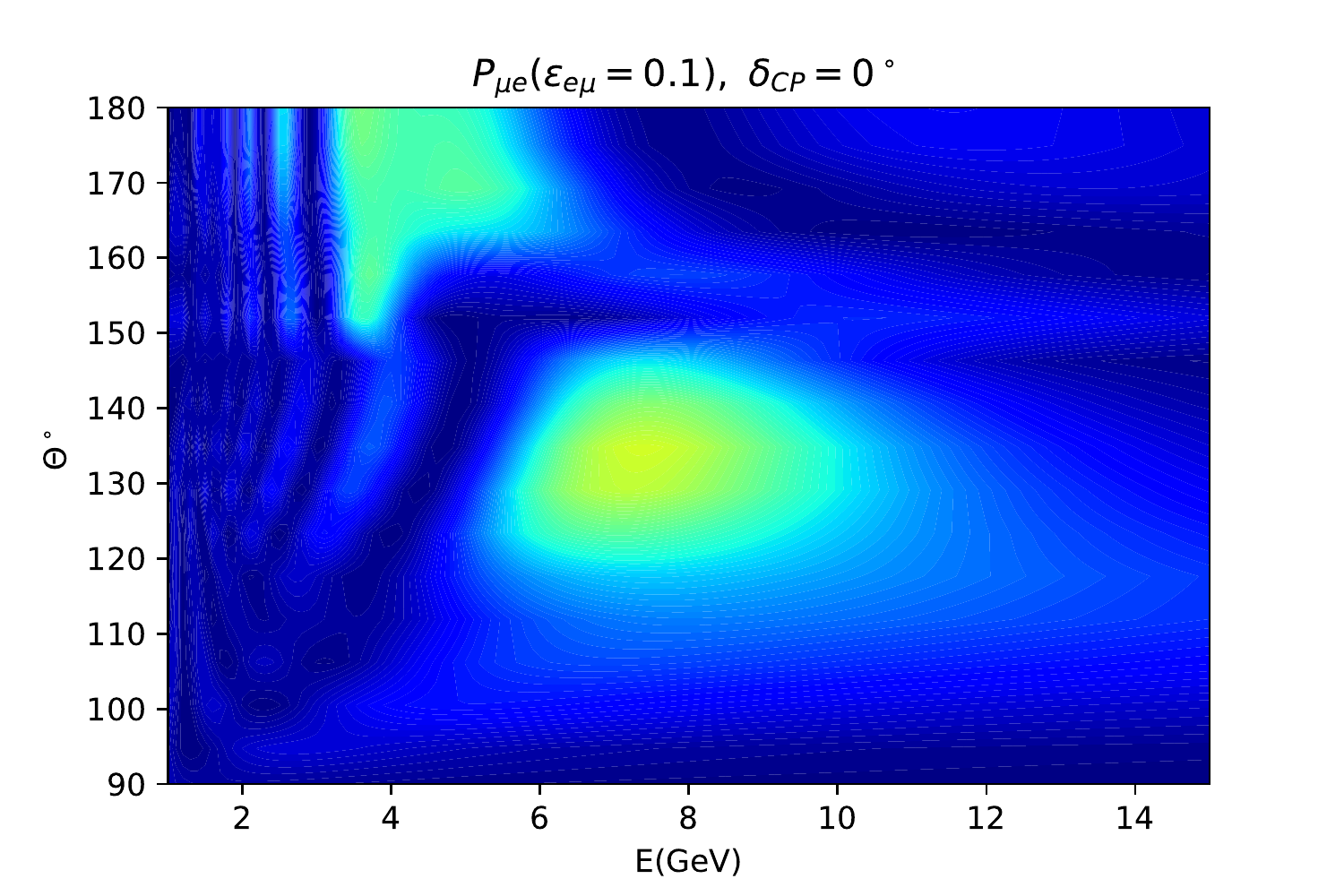}
\includegraphics[width=0.44\textwidth]{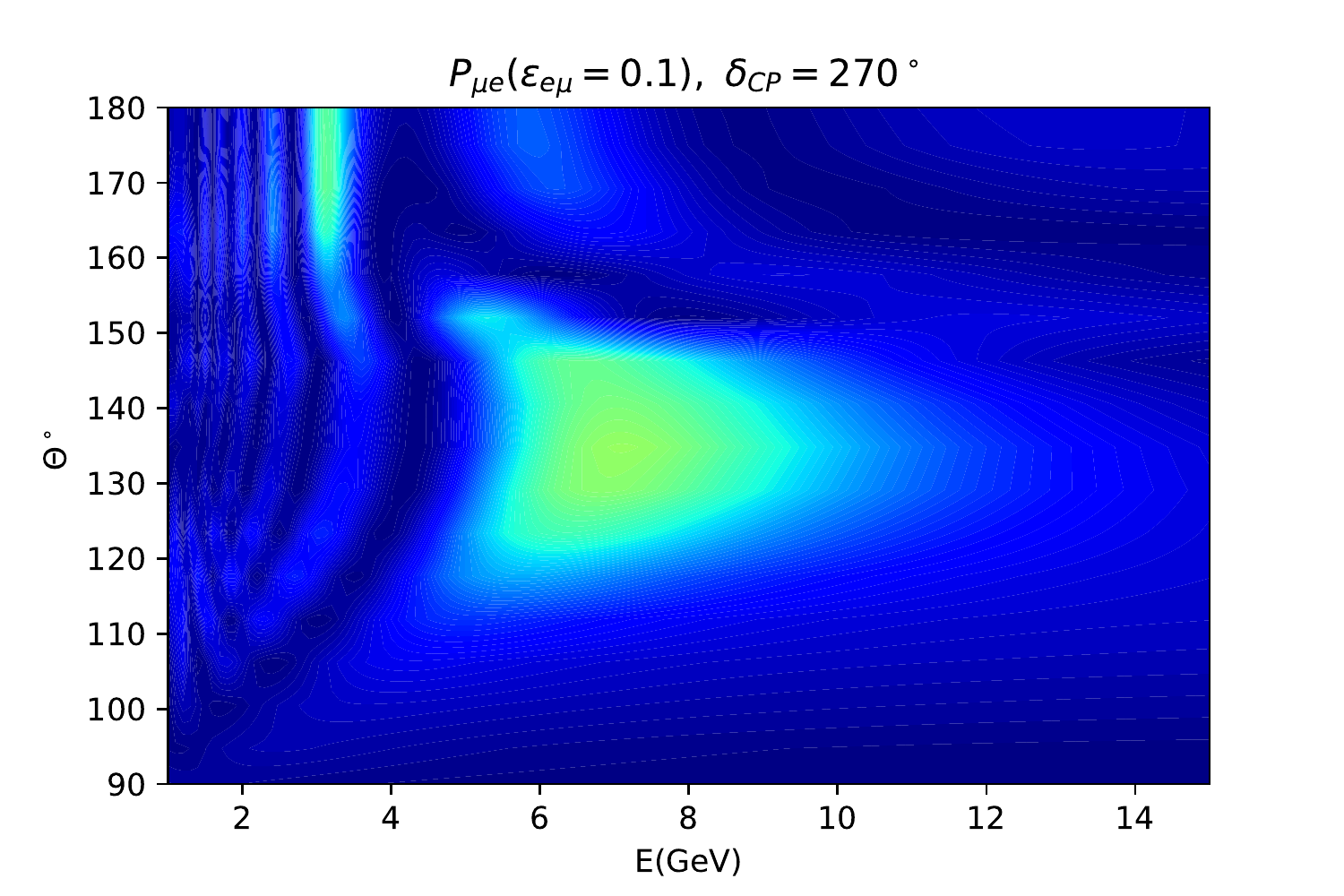}
\end{center}
\caption{\label{cor44}    Oscillograms for the probability $P_{ \mu e }$  assuming $\delta_{CP}=0^\circ$ (left panels) and $\delta_{CP}=270^\circ$ 
{\color{blue}}
(right panels). The upper panels indicate the oscillogram for the case of no NSI.  The middle left (right) panels show the oscillogram for the case of   non-zero $| \varepsilon_{e e} - \varepsilon_{\mu \mu} |$  while setting other NSI parameters to zero. The lower panels are plotted assuming non-zero $\varepsilon_{e\mu }$. Normal neutrino mass hierarchy is assumed to be
true and the oscillation parameters are set as given in \cite{Esteban:2020cvm}.  
}
\end{figure}

On the other hand, the oscillation probabilities can depend on the values of $\delta_{CP}$ in the low energy range of $1-15$ GeV, due to the $\theta_{13}$ resonance effect, as shown in Fig.~\ref{deltaa}.
Moreover, there are correlations between the NSI parameters ($\varepsilon_{e \mu}$, $\varepsilon_{e \tau}$ and $| \varepsilon_{ e e } - \epsilon_{\mu \mu} |$) and $\delta_{CP}$ which can affect the oscillation probabilities. 

The neutrino oscillograms of $P_{\mu e}$ without and with the NSIs for the low energy region $E \in [1, 15]$ GeV are shown in Fig.~\ref{cor44}.
As a reference, we compare the cases where $\delta_{CP} = 0^\circ$ (no CP violation) and $270^\circ$ (maximal CP violation) since the recent analyses claim the best CP violation phase is around $270^\circ$~\cite{Esteban:2019lfo,Esteban:2020cvm}.
As we observe, these plots show the correlations between $\delta_{CP}$ and the NSI parameters $\varepsilon_{e \mu} $ and $ | \varepsilon_{e e} - \varepsilon_{\mu \mu} |$. 
For $\varepsilon_{e \tau}$ which we have not shown here, the result is similar to the $\varepsilon_{e \mu}$ case.

The oscillation probability of $\nu_\mu \to \nu_e$    for $E \gtrsim 15$ GeV, is almost independent of the Zenith angle, $\delta_{CP}$, and the NSI parameter $|\varepsilon_{ee} - \varepsilon_{\mu \mu}|$; on the other hand, $P_{\mu e}$ moderately decreases in the presence of $\varepsilon_{e\mu}$.
Note that the color codes are the same as those in Fig.~\ref{oscmm} and \ref{oscmt}.
For lower energy region $E \lesssim 10$ GeV, we see the sensitivity of $P_{\mu e}$ on $\varepsilon_{e\mu}$ is notable by comparing the top-left and the bottom-left panels.
The sensitivity on $\delta_{CP}$ remarkably increases in the presence of the non-zero NSI parameter $\varepsilon_{e\mu}$, as depicted in the bottom panels.
On the other hand, $P_{\mu e}$ is not  sensitive on the other parameter $|\varepsilon_{ee} - \varepsilon_{\mu \mu}|$ although one can see slight changes for large Zenith angle $\Theta \gtrsim 150^\circ$.
We confirmed the oscillograms of $\nu_e \to \nu_\mu$ show almost the same behavior as  Fig.~\ref{cor44} and we do not show them in this paper.

\section{Results}
\label{sec:results}

In this paper, we consider the ten years of future data taking of the reference experiments: DUNE, HK, KNO, and ORCA.
We take the detailed information of the DUNE far detectors in Refs.~\cite{ DUNE:2020ypp} based on the proposal of installing the four liquid argon time projection chambers (LArTPCs) with a fiducial volume of 40 kt in total.
The information of the HK detector is taken from Ref.~\cite{Hyper-Kamiokande:2018ofw} installing a single water Cherenkov light detector with a fiducial volume of 260 kt. 
The KNO experiment which aims to combine its astrophysical neutrino observations and the beam neutrino detection coming from the J-PARC in close collaboration with HK is expected to be installed in the south-east part of Korea and take data from 2027~\cite{kno}.
Although not definitely decided, we conservatively assume the fiducial volume of the water Cherenkov light detector is 260 kt and the detection technology is the same as that of HK for simplicity so that they can combine the data easily.
The ORCA is expected to have a fiducial volume of 6 Mt with an energy threshold of a few GeV.
Here we assume the energy threshold is 3 GeV; the detection efficiency is assumed to be 50\% for $3\,{\rm GeV} \le E \le 10\,{\rm GeV}$ and 90\% for $E > 10$ GeV~\cite{KM3Net:2016zxf}.
The detailed information of fiducial volume, energy threshold of detection of the muon/electron neutrino $E_{\rm th}$, and the $\nu_\mu$ detection efficiency of each experiment is summarized in Table~\ref{tab:experiments}.
\begin{table}[h]
    \centering
    \begin{tabular}{|c|c|c|c|c|}
    \hline
        Experiments & ~~ $E_{\rm th}$ for $\nu_{\mu}$ ~~ &
         ~~ $E_{\rm th}$ for $ \nu_{e}$ ~~ & fiducial volume &  $\nu_e $ and $\nu_\mu $ detection efficiency   \\
        \hline
        DUNE & $ 135 ~ {\rm MeV}$ & $ 10 ~ {\rm MeV}$     &  40 kt & $85 \%$ \\
        \hline
        HK  & $ 110 ~ {\rm MeV}$ & $ 6.5 ~ {\rm MeV}$  & 260 kt & $80 \%$  \\
        \hline
        KNO & $ 110 ~ {\rm MeV}$ & $ 6.5 ~ {\rm MeV}$ & 260 kt & $80 \%$ \\
        \hline
        ORCA & $3 ~ {\rm GeV}$ & $3 ~ {\rm GeV}$ & 6 Mton & ~ $50 \%$ ($ E<10~ {\rm GeV}$)~,~ $90 \%$  ($E>10~ {\rm GeV}$) ~ \\
        \hline
    \end{tabular}
    \caption{The detailed information of the reference experiments for this analysis \cite{DUNE:2020ypp,Kelly:2021jfs, KM3Net:2016zxf,Hyper-Kamiokande:2018ofw}.}
    \label{tab:experiments}
\end{table}

For a given ordering (normal ordering or inverted ordering) and best-fit values of the oscillation parameters, the oscillation probability depends only on the neutrino energy and the zenith angle of the direction of the neutrino.
We assume the mass ordering is known to be normal from the precise measurements of JUNO and RENO-50 in this paper~\cite{Bakhti:2020fde}.
Notice that JUNO and RENO-50 experiments are not sensitive to NSI due to their low neutrino energy. 
In the following, we set the true values of the oscillation parameters as the best-fitted points given in Ref.~\cite{Esteban:2020cvm}, except $\delta_{CP} $. 
We have taken the flux of atmospheric neutrinos and the cross-section from Refs.~\cite{Honda:2006qj} and \cite{Kajita:2000mr}, respectively.

For the statistical inferences, we have considered the Asimov data set approximation.
We have assumed the standard paradigm (no NSI) as the true model and quantified the difference between the standard interaction (SI) and NSI events in terms of the chi-squared function defined as
\begin{align}
\chi^2 = \min \sum_{ij}\left[\frac{({N }^{\rm th}_{ij} - N^{\rm ex}_{ij})^2}{N^{\rm ex}_{ij}}\right]
+ \sum_{s=1}^k\xi_s^2 \,,
\end{align}
where ${N}^{\rm th}_{ij} $ and $N^{\rm ex}_{ij}$ are the predicted and expected numbers of events in the $i$th zenith angle bin and the $j$th energy bin, respectively. 
According to our set-up, $N_{ij}^{\rm ex} = N_{ij}^{\rm th} (\varepsilon = 0)$, i.e., the theoretical number of events in the absence of NSIs. 
The predicted number of events $N_{ij}^{\rm th}$ of a neutrino $\nu_\alpha$ in each bin is theoretically calculated as
\begin{align}
    N_{ij}^{\rm th}(\epsilon)=TN\int d\Omega \int dE_\nu \int  \sigma(E_\nu) [\phi_\mu(E_\nu) P_{\mu\alpha}(E_\nu,\varepsilon,\Omega)+\phi_e(E_\nu) P_{e\alpha}(E_\nu,\varepsilon,\Omega) ] R_{ij}(E_\nu)    \,,
\end{align}
where $T$ is the exposure time, $N$ is the number of target particles, $\Omega$ is the solid angle, $\phi_\beta(E_\nu)$ is the neutrino flux of flavour $\beta$ at detector, $\sigma(E_\nu)$ is detection cross section of neutrino $\alpha$, and $P_{\beta\alpha}$ is the oscillation probability from flavor $\beta$ to flavor $\alpha$. 
Note that we included a detector response function $R_{ij}(E_\nu)$ for which Gaussian resolution function is a good approximation and is given as
\begin{align}
R_{ij}(E_\nu)=\frac{1}{2}\left[ erf\left( \frac{E^{up}_j-E^{true}}{\sqrt{2}\sigma_{E_\nu}}\right)- erf\left( \frac{E^{true}-E^{down}_j}{\sqrt{2}\sigma_{E_\nu}}\right)\right]\,,    
\end{align}
where $\sigma_{E_\nu}$ is the energy resolution, $E^{down}_j$ is the lower limit of energy of the $j$-th bin, $E^{up}_j$ is the upper limit of energy of the $j$-th bin, and $E^{true}$ is the real energy of the neutrino. 
We find our total number of events ($\sum_{ij}  N_{ij}^{\rm th}$) value in good agreement with those in the technical or conceptual design reports of the reference experiments.

The corresponding nuisance parameters or pull variables \{$\xi_s\}$ include the theoretical and systematic uncertainties in terms of $k$ independent sources of error~\cite{Fogli:2002pt,Behera:2016kwr}.
In the performed analysis we considered one $\xi_s$ to be the normalization uncertainty of 20\% in each experiment.
Since the number of our reference experiments is four, we have $k=4$.

To perform our analysis, we have taken three equal size zenith angle bins in $ \Theta $ between $90$ to $180$ degrees, and twenty equal size energy bins, between 1~GeV and 200~GeV.
We split our analysis according to the energy range: $E > 15$ GeV, dubbed {\it high energy range}, and $E \le 15$ GeV, dubbed {\it low energy range}.
As seen in the previous section, we can effectively consider the two flavor approximation assuming $P_{ee} \sim 1$ in the high energy range and hence the relevant NSI parameters are just $\varepsilon_{\mu \tau}$ and 
$|\varepsilon_{\mu \mu} - \varepsilon_{\tau \tau}|$.
We consider $\nu_\mu$/$\bar \nu_\mu$ detection (disappearance) in all the reference experiments and add $\nu_\tau$/$\bar \nu_\tau$ detection (appearance) in DUNE.
In the low energy range, the two flavor approximation does not work due to the $\theta_{13}$ resonance effect, as stated in the previous section.
Hence we can observe the correlations between the CP violation phase $\delta_{CP}$ and the NSI parameters ($\varepsilon_{e \mu}$, $\varepsilon_{e \tau}$ and $| \varepsilon_{ e e } - \epsilon_{\mu \mu} |$).

\subsection{High energy range results}
\label{sec:results-HE}

\begin{figure}
\begin{center}
\includegraphics[width=0.49\textwidth]{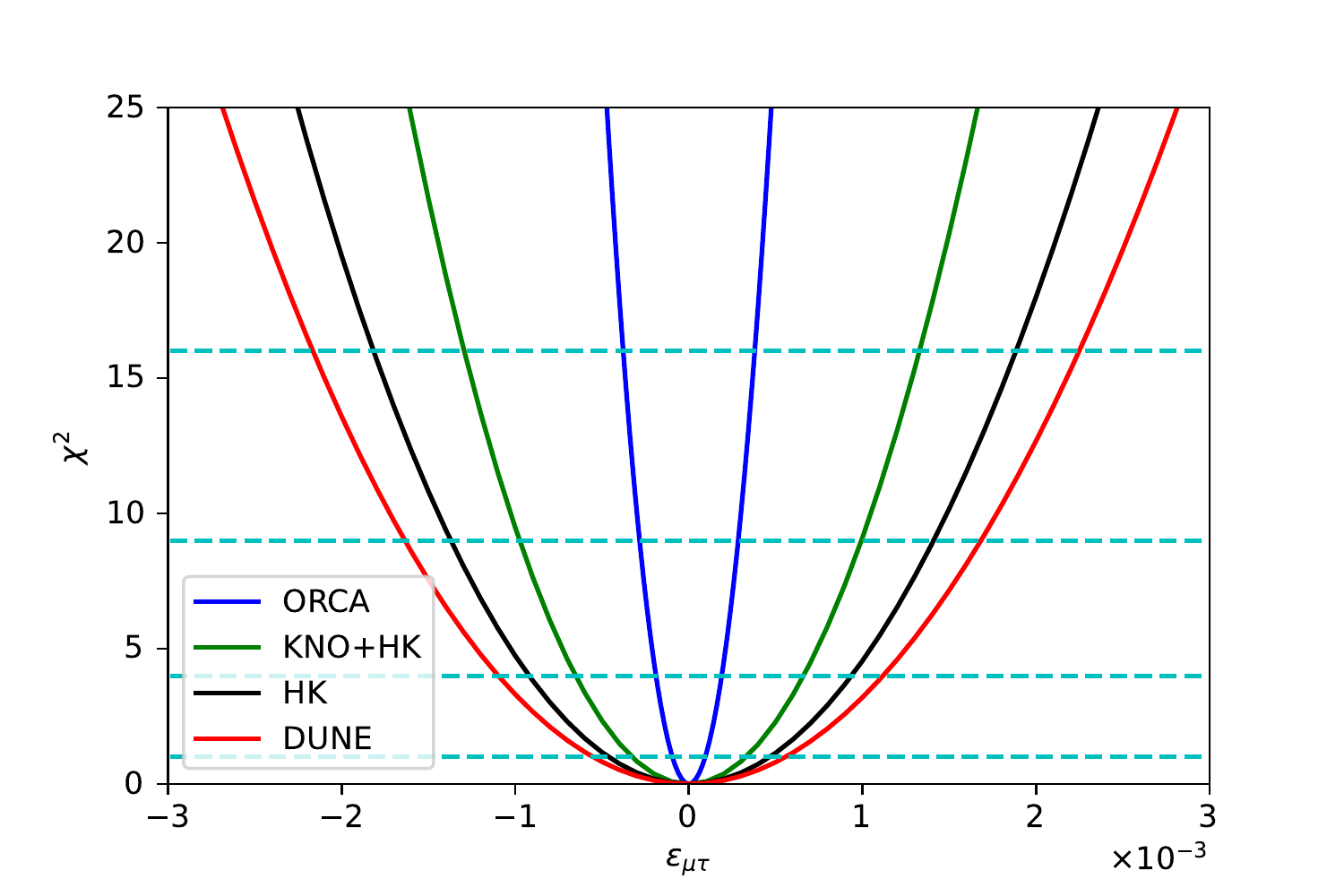}
\includegraphics[width=0.49\textwidth]{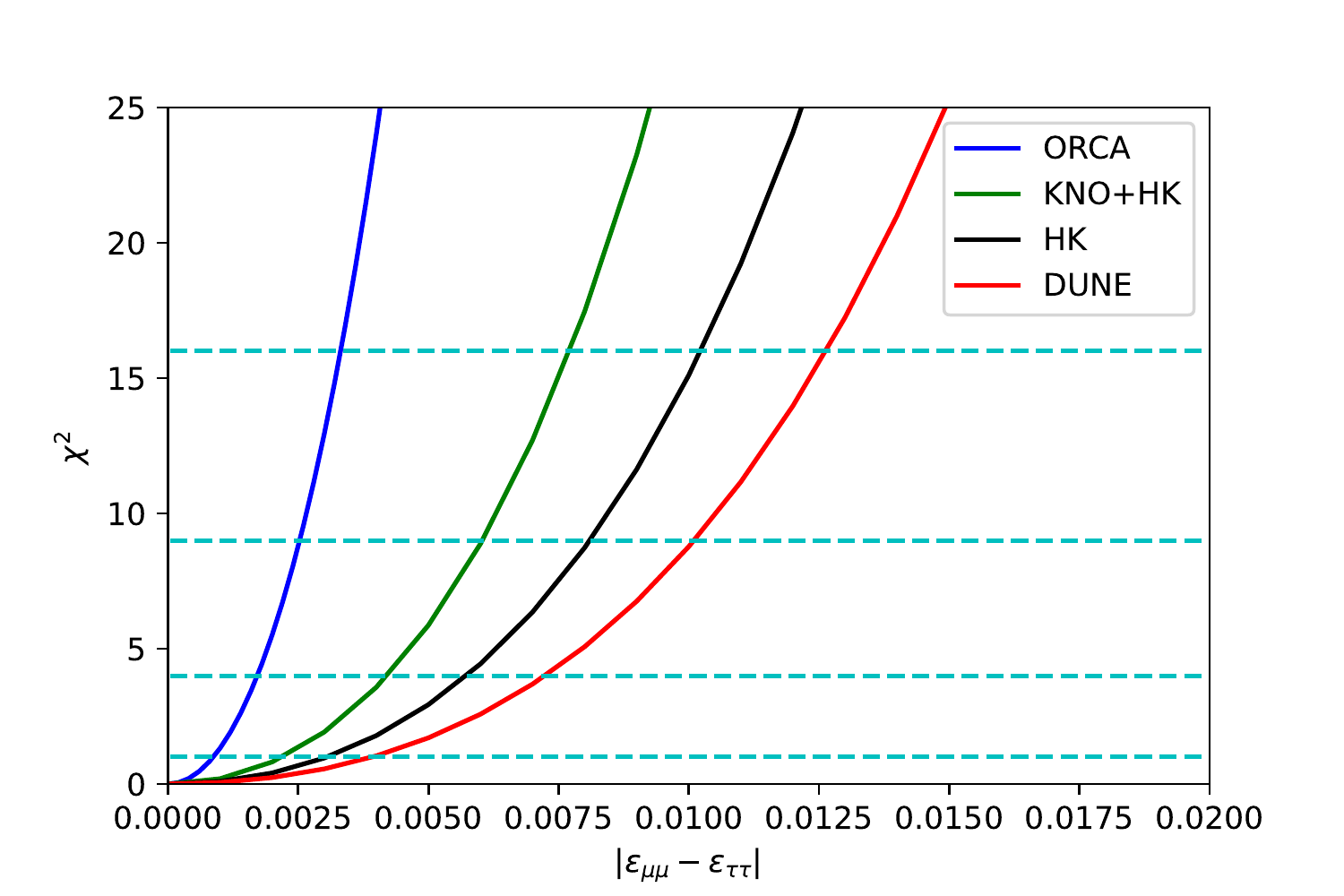}
\end{center}
\caption{\label{chi22} The expected Chi-squared
for NSI as a function of $\varepsilon_{\mu \tau}$ (right) and $ | \varepsilon_{\mu \mu} - \varepsilon_{\tau \tau} |$ (left) assuming the 10 years of data taking by DUNE (red curve), HK (black curve), KNO combined with HK (green curve), and ORCA (blue curve). For DUNE, We have assumed the detection efficiency of $\nu_ \tau$ as $30\%$.
The neutrino mass hierarchy is assumed to be known and normal. The horizontal dashed green lines show the $1\sigma$, $2\sigma$, $3\sigma$ and $4\sigma$ lines.
}
\end{figure}

Equipped with the statistical analysis method, we now investigate the potential of our reference experiments in probing the NSI parameters $\varepsilon_{\mu \tau} $ and $| \varepsilon_{\mu \mu} - \varepsilon_{\tau \tau} |$ in the high energy range $E > 15$ GeV.
Due to the ability of detecting $\nu_\tau$ event by event in DUNE~\cite{Conrad:2010mh}, we include the $\nu_\tau$ appearance in the experiment assuming the $\nu_\tau$ detection efficiency of 30\% in all energy ranges we consider~\cite{DUNE:2020ypp}.
Note that this is an optimistic value and a more realistic value should be studied further~\cite{DeGouvea:2019kea,nutauDUNE}.
A more optimistic claim from Ref.~\cite{DUNE:2020ypp} is that the $\nu_\tau$ detection efficiency can be near to 100\% for high energies.
We show how this optimistic scenario can change the sensitivities of DUNE in the Appendix.

The other reference experiments can detect $\nu_\tau$ only statistically, inferred from the oscillation of $\nu_\mu \to \nu_\tau$ (not by events) and hence we do not include the $\nu_\tau$ detection there.
The sensitivities in terms of the $\chi^2$ by taking $\varepsilon_{\mu \tau}$ ($| \varepsilon_{\mu \mu} - \varepsilon_{\tau \tau} |$) the only free NSI parameter while setting the others to zero are shown in the left (right) panel of Fig.~\ref{chi22}.
The solid red, black, green, and blue curves correspond to the expected sensitivities after the 10 years of data taking from DUNE, HK, KNO combined with HK (denoted as KNO+HK), and ORCA detectors, respectively. 
The horizontal dashed cyan lines correspond to the $1\sigma$, $2\sigma$, $3\sigma$ and $4\sigma$ lines.
As can be observed, the size of the fiducial volume plays a key role in increasing the sensitivities of the NSI parameters $ \varepsilon_{\mu \tau} $ and $| \varepsilon_{\mu \mu} - \varepsilon_{\tau \tau} |$; ORCA has the best sensitivities with the highest statistics.
Note that the sensitivity of DUNE on $\varepsilon_{\mu \tau}$ is nevertheless close to that of HK due to the inclusion of the $\nu_\tau$ appearance with the detection efficiency of 30\% in spite of its much smaller fiducial volume, which proves the effectiveness of detecting the $\tau$ neutrino appearance event by event.
Therefore, increasing the $\nu_\tau$ detection efficiency would play an important role in making DUNE competitive over the other larger size experiments.~\footnote{See Ref.~\cite{Machado:2020yxl} for a recent effort to increase the $\nu_\tau$ detectability by applying the methods in collider physics, focused on the beam-induced neutrinos.}
For $|\varepsilon_{\mu \mu} - \varepsilon_{\tau \tau}|$, however, the inclusion of the $\nu_{\tau}$ detection channels does not improve the sensitivity.
The situation is the same even when the $\nu_\tau$ detection efficiency is 100\% as shown in Fig.~\ref{chi11} in Appendix.
It is also remarkable that the combination of KNO with HK shows high sensitivities due to the huge fiducial volume.

The future $1\sigma$ sensitivities on the NSI parameters in Fig.~\ref{chi22} are summarized as follows.
\begin{align}
{\rm ORCA}&:~~~~~~~\varepsilon_{\mu \tau} < 1.0 \times 10^{-4} ~~~,~~~~| \varepsilon_{\mu \mu} - \varepsilon_{\tau \tau} | < 9 \times 10^{-4}~~, \\
{\rm KNO + HK}&:~~~~~~~\varepsilon_{\mu \tau} < 2.8 \times 10^{-4} ~~~,~~~~| \varepsilon_{\mu \mu} - \varepsilon_{\tau \tau} | < 1.9 \times 10^{-3}~~, \\
{\rm HK}&:~~~~~~~\varepsilon_{\mu \tau} < 4.0 \times 10^{-4} ~~~,~~~~| \varepsilon_{\mu \mu} - \varepsilon_{\tau \tau} | < 2.6 \times 10^{-3}~~, \\
{\rm DUNE}&:~~~~~~~\varepsilon_{\mu \tau} < 5.5 \times 10^{-4} ~~~,~~~~| \varepsilon_{\mu \mu} - \varepsilon_{\tau \tau} | < 3.9 \times 10^{-3}~~.
\end{align}
It is notable that the experimental sensitivities can be significantly improved by around two orders of magnitude from the current bounds summarized in Ref.~\cite{Esteban:2019lfo}.

\subsection{Low energy range results}
\label{sec:results-low}

\begin{figure}
\begin{center}
        \includegraphics[width=0.49\textwidth]{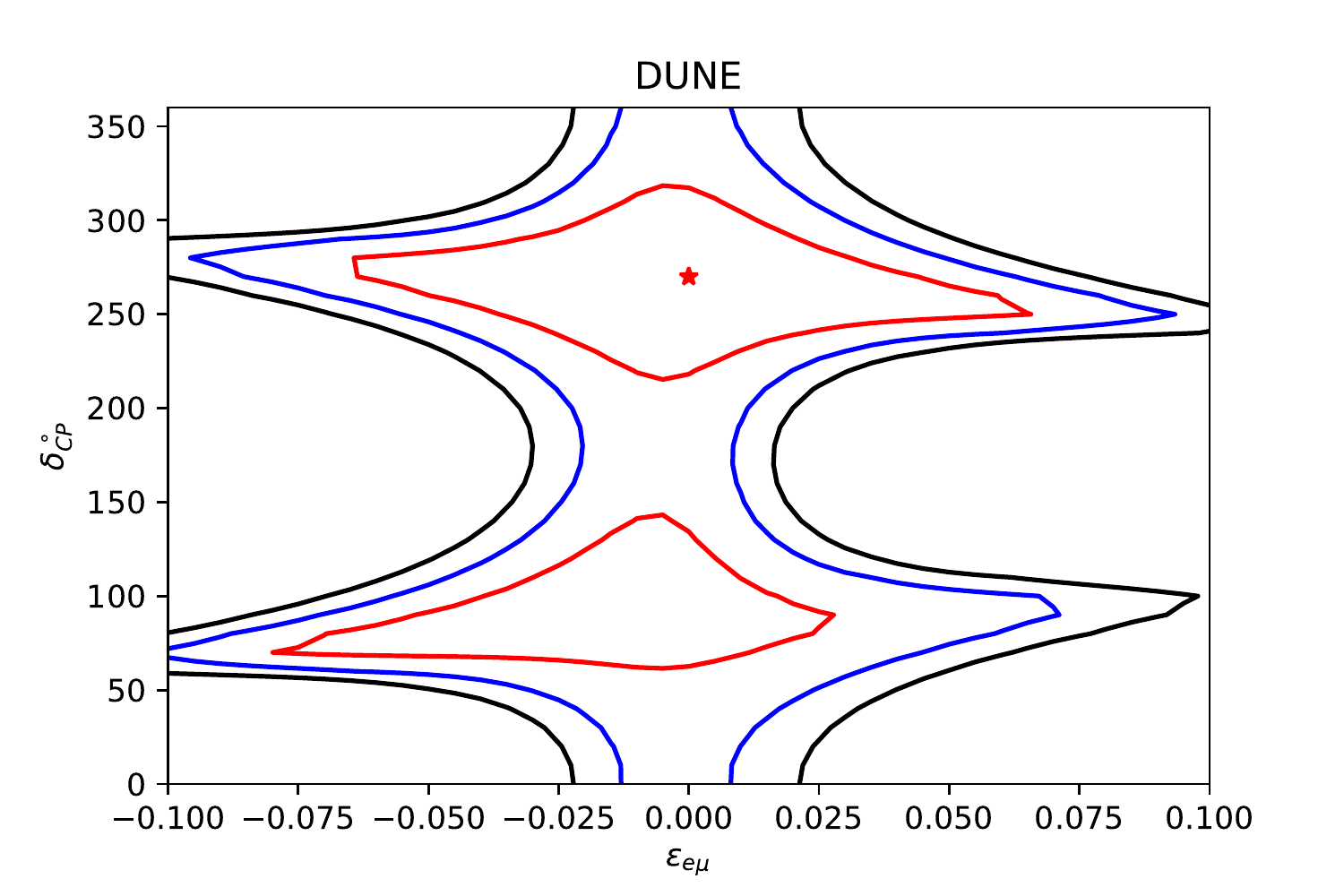}
        \includegraphics[width=0.49\textwidth]{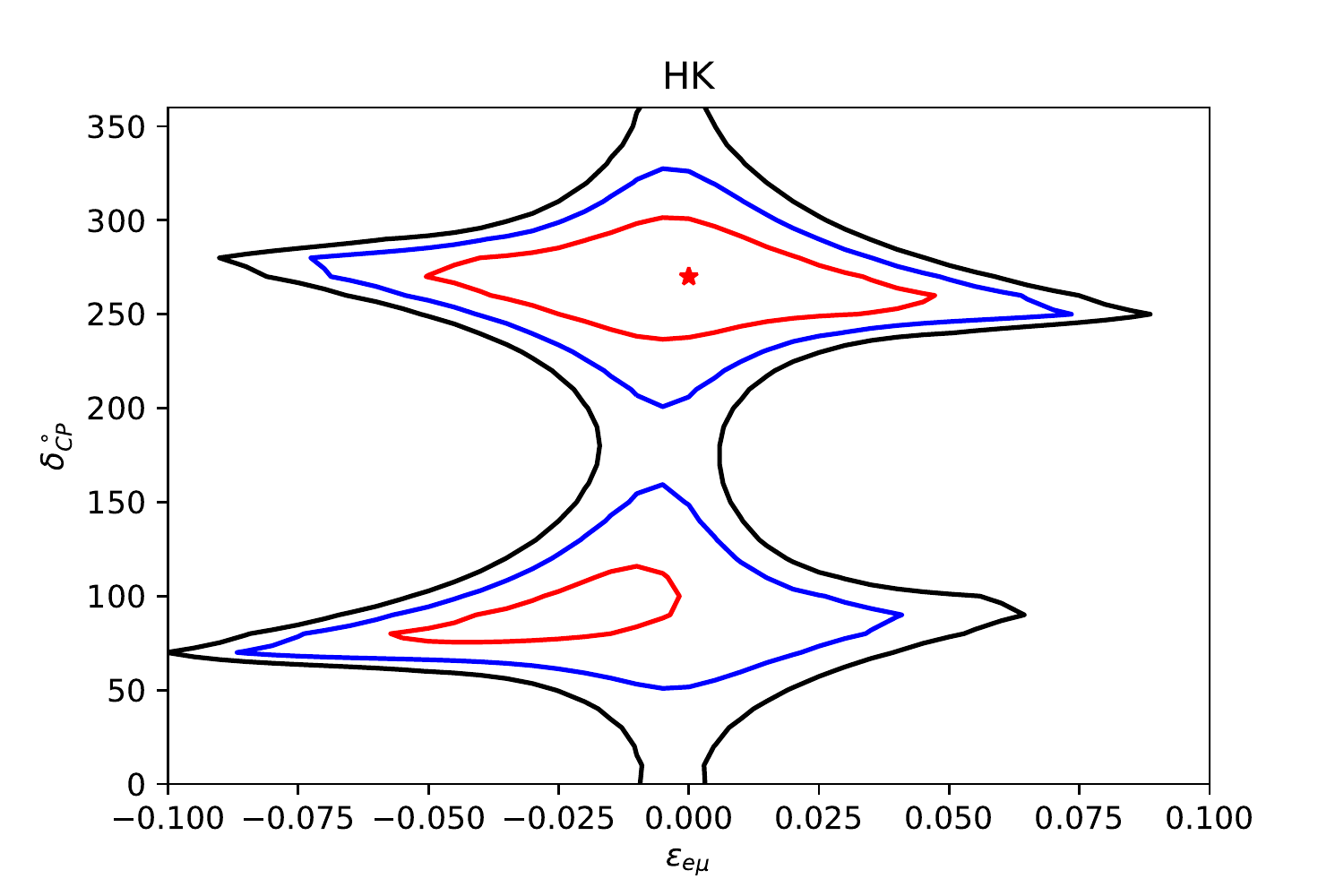}
        \includegraphics[width=0.49\textwidth]{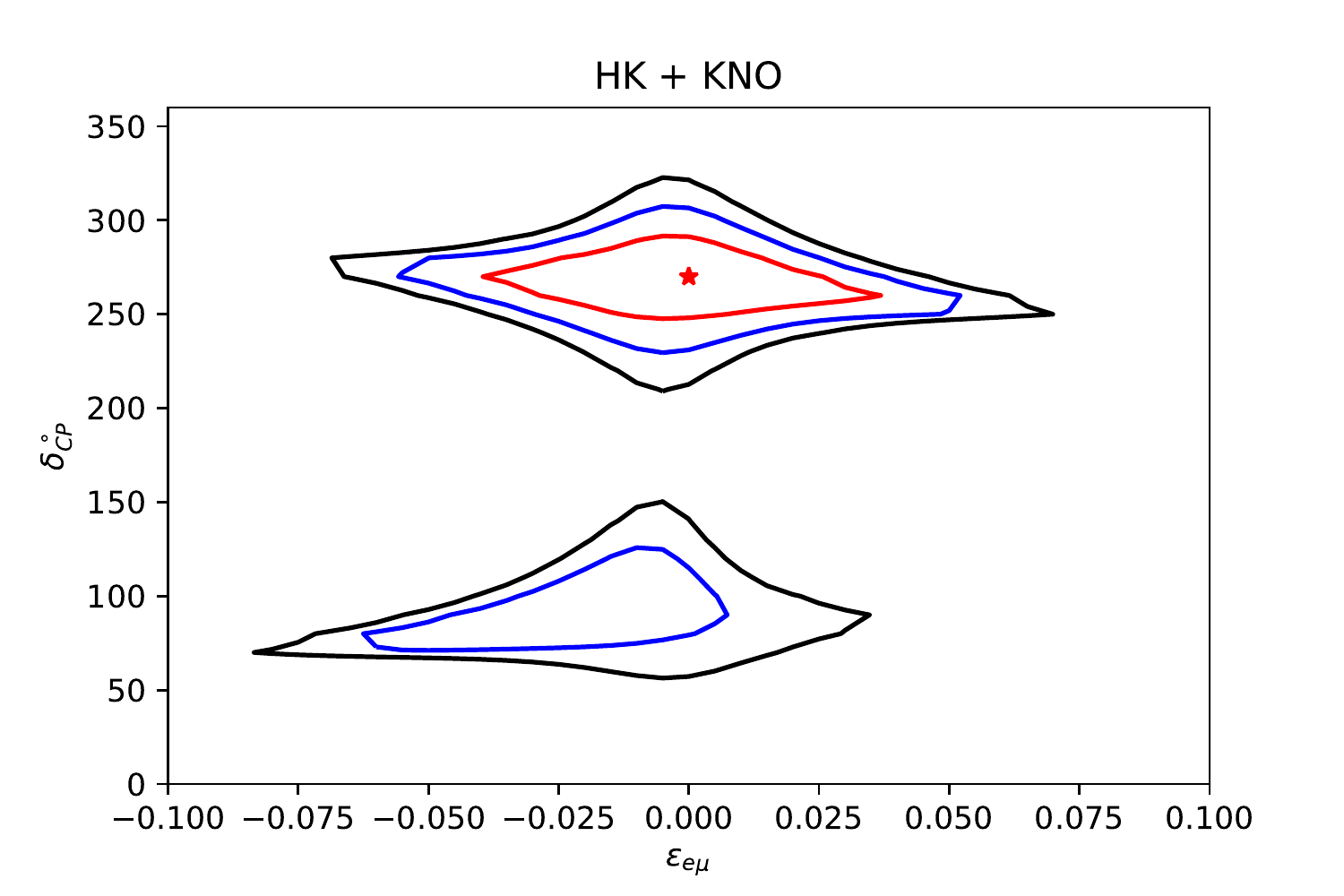}
        \includegraphics[width=0.49\textwidth]{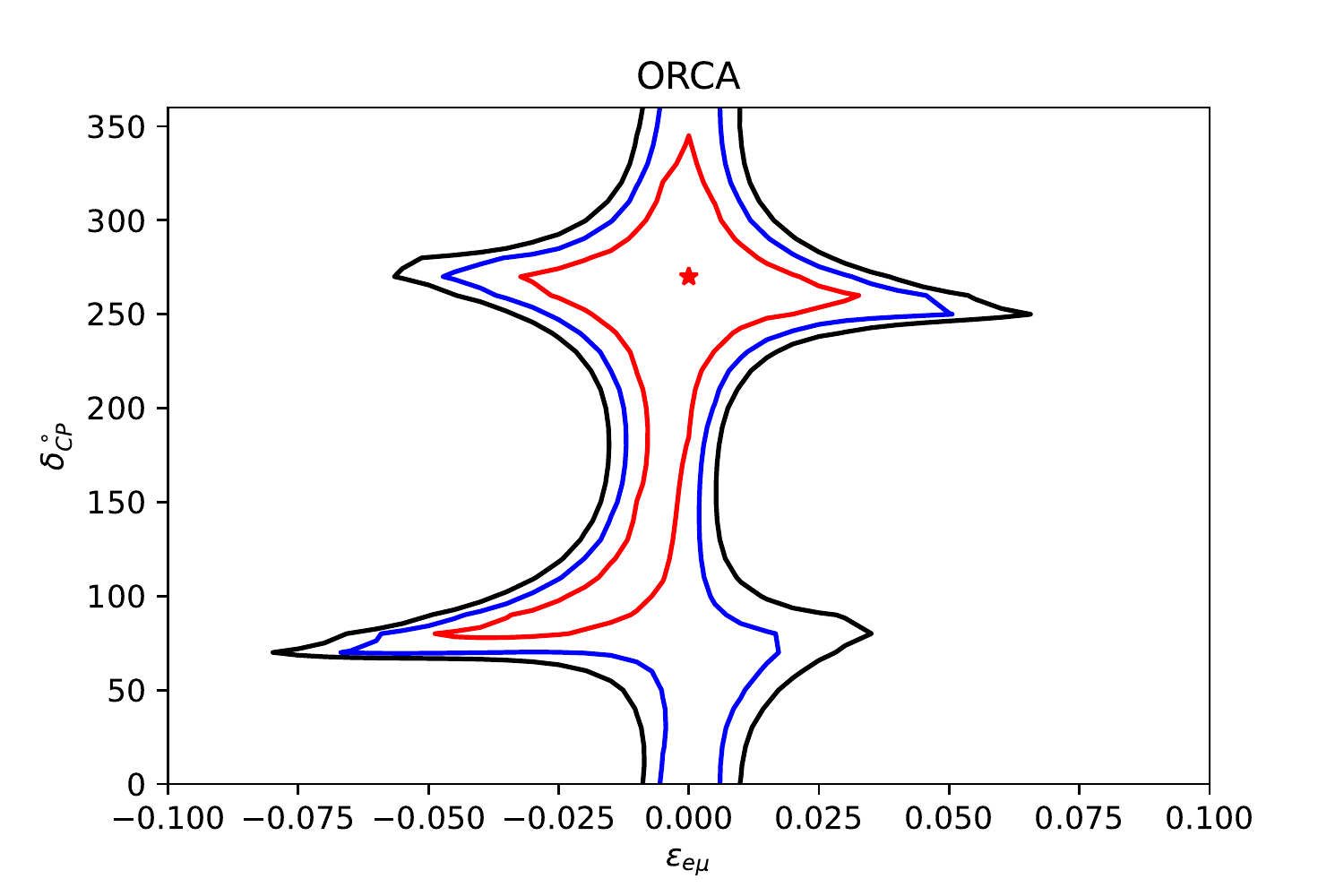}                
 \end{center}
\caption{\label{cord} $1\sigma$, $2\sigma$ and $3\sigma$ C.L. contours in the $\delta_{CP} - \varepsilon_{e \mu}$ plane    expected from 10  years of running of the DUNE (upper left), HK (upper right), KNO $+$ HK (lower left) and ORCA (lower right) experiments. We have assumed the true value of $\delta_{CP} = 270^{\circ} $. Known normal neutrino mass hierarchy is assumed to be
true. We have assumed the standard model as the true model.}
\end{figure}

\begin{figure}
\begin{center}
        \includegraphics[width=0.49\textwidth]{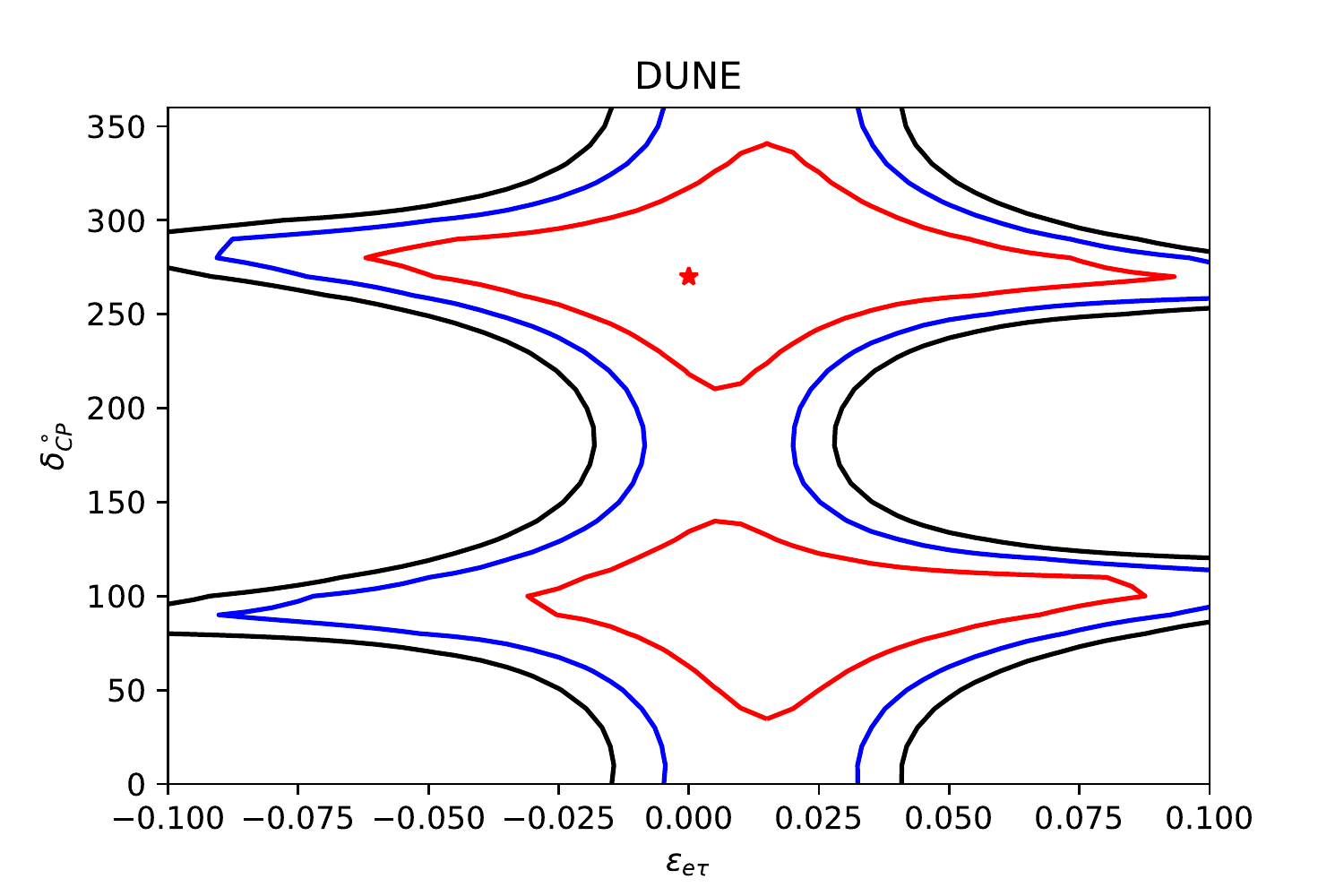}
        \includegraphics[width=0.49\textwidth]{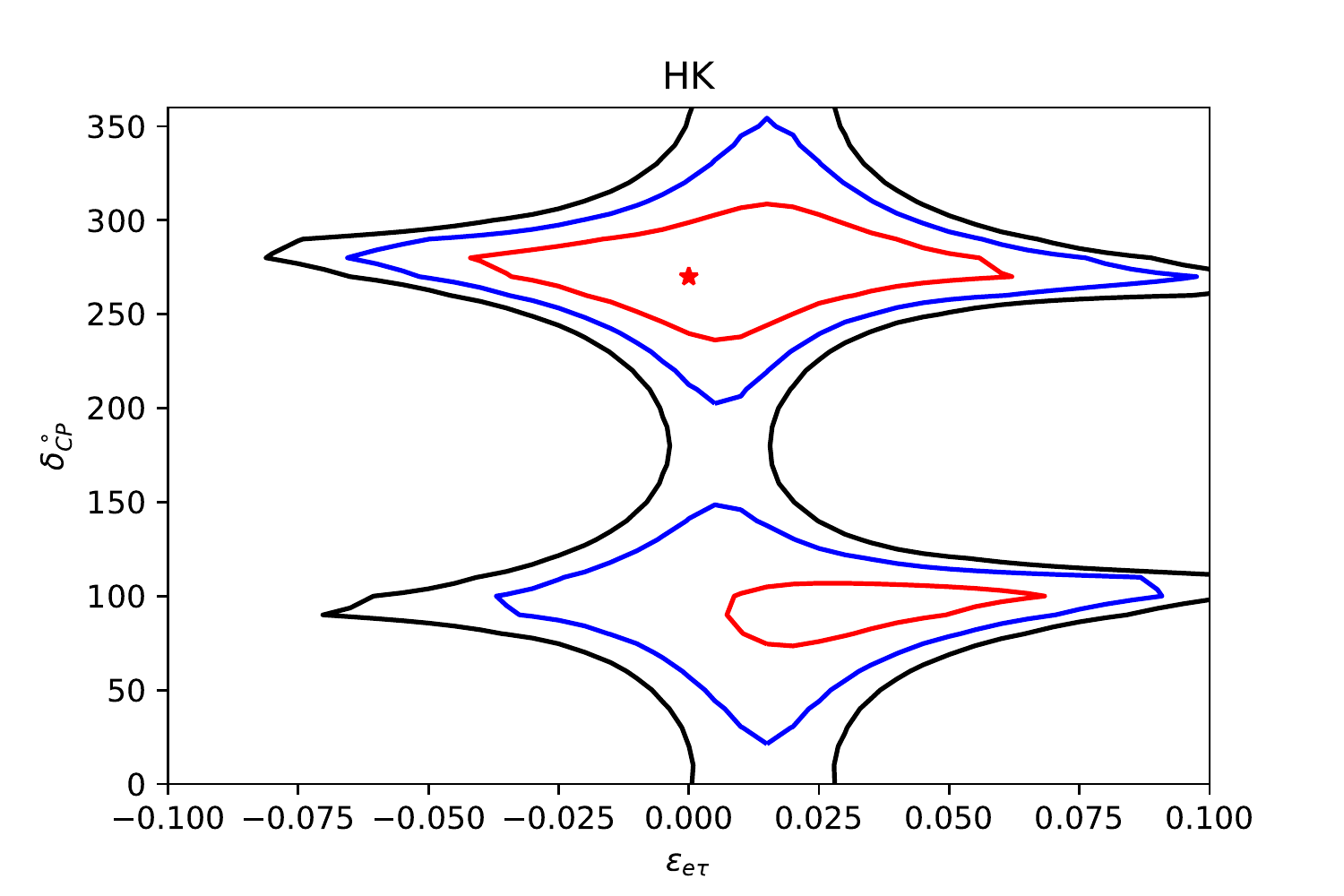}
        \includegraphics[width=0.49\textwidth]{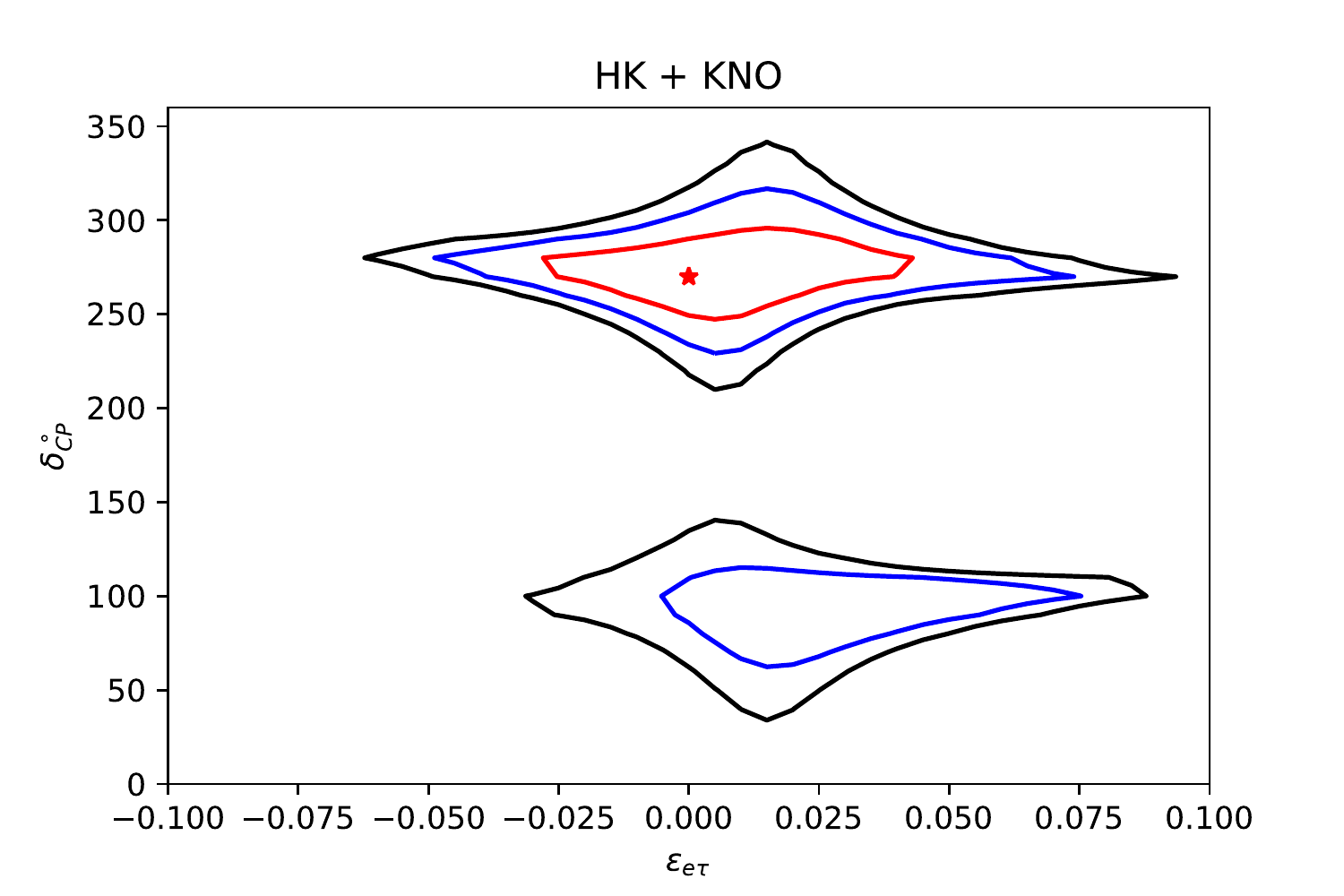}
        \includegraphics[width=0.49\textwidth]{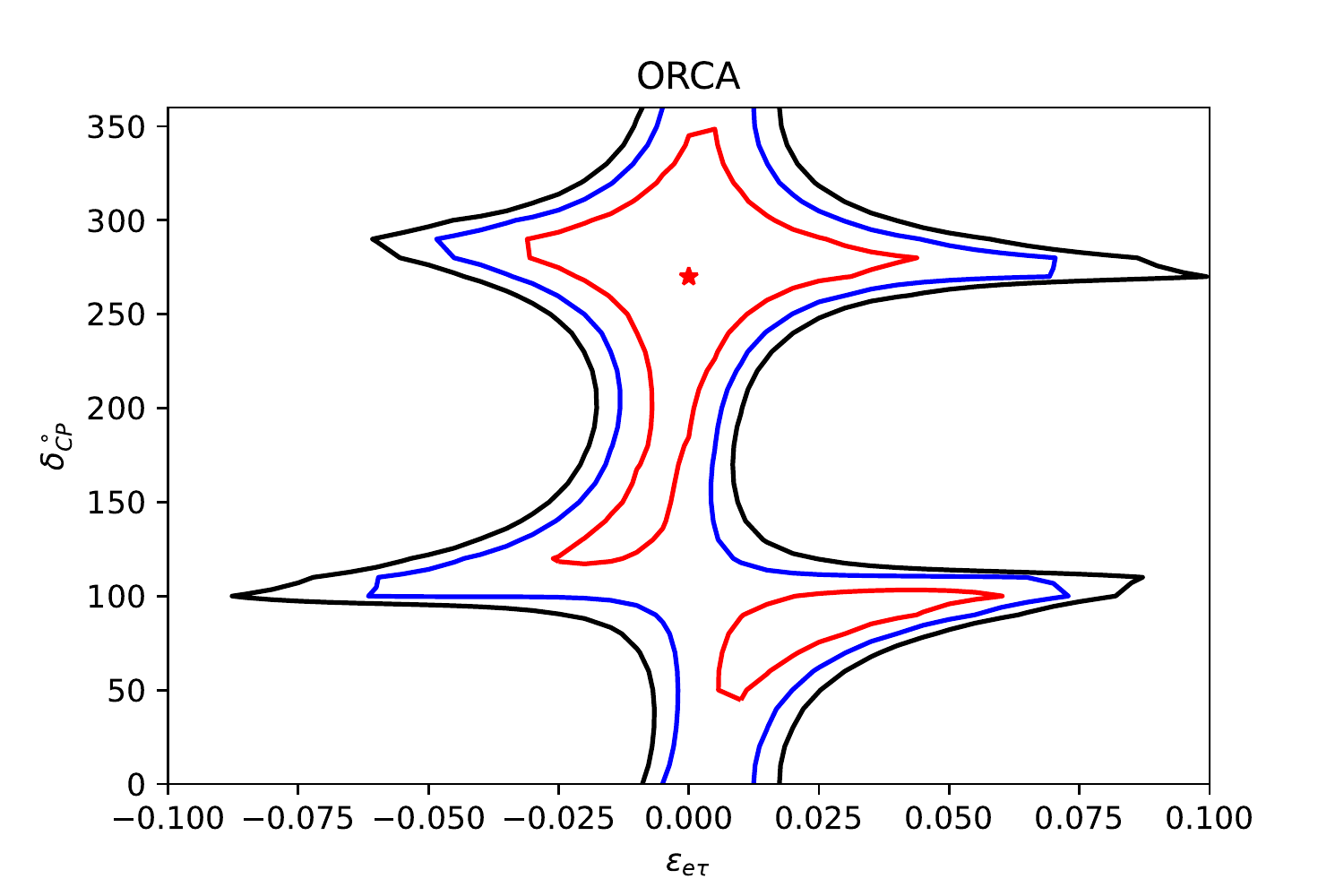} 
 \end{center}
\caption{\label{cord2} $1\sigma$, $2\sigma$ and $3\sigma$ C.L. contours in the $\delta_{CP} - \varepsilon_{e \tau}$    expected from 10  years of running of DUNE (upper left), HK (upper right), KNO $+$ HK (lower left) and ORCA (lower right)  experiments. We have assumed the true value of $\delta_{CP} = 270^{\circ} $. Known normal neutrino mass hierarchy is assumed to be
true. We have assumed the standard model as the true model. }
\end{figure}

\begin{figure}
\begin{center}
        \includegraphics[width=0.49\textwidth]{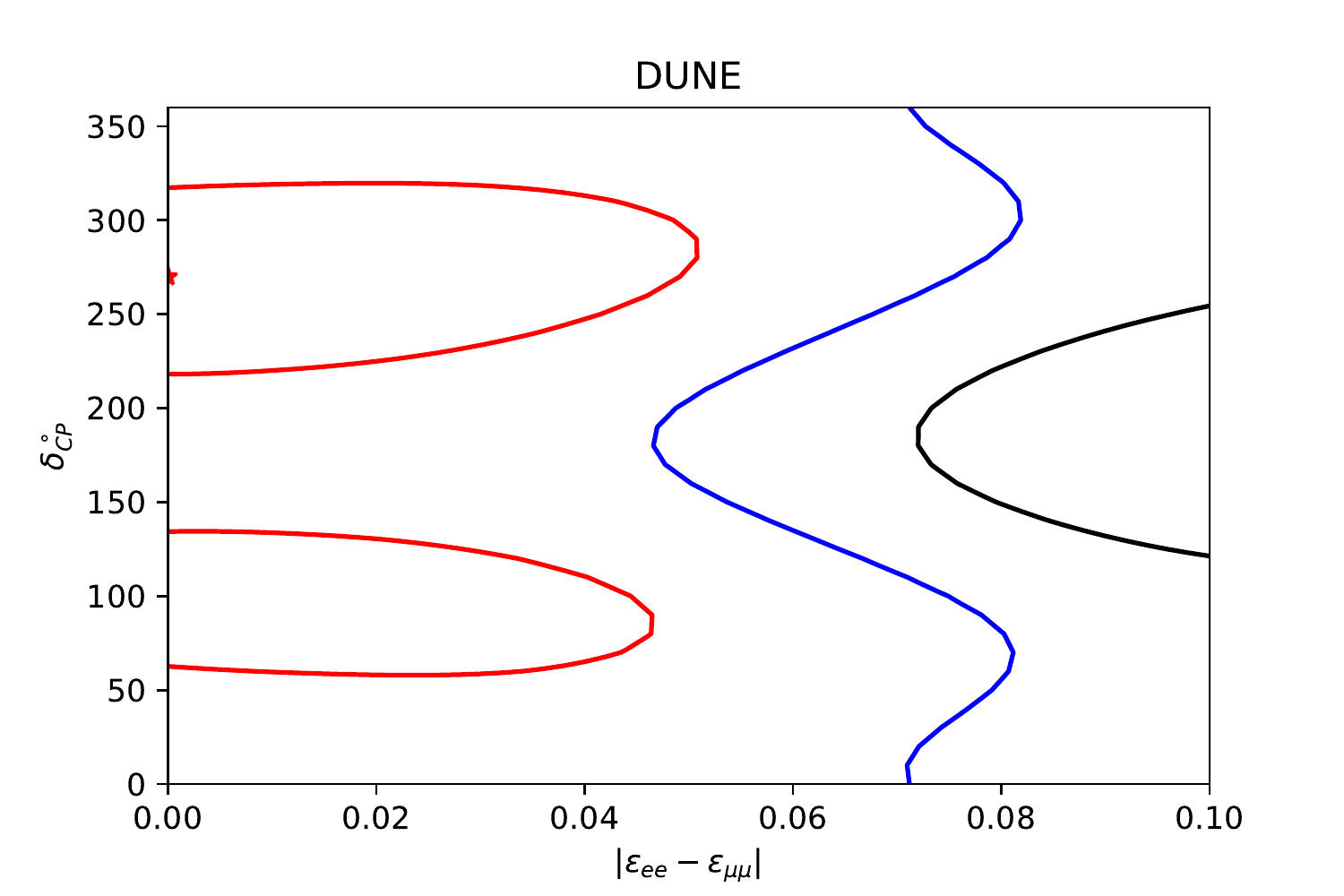}
        \includegraphics[width=0.49\textwidth]{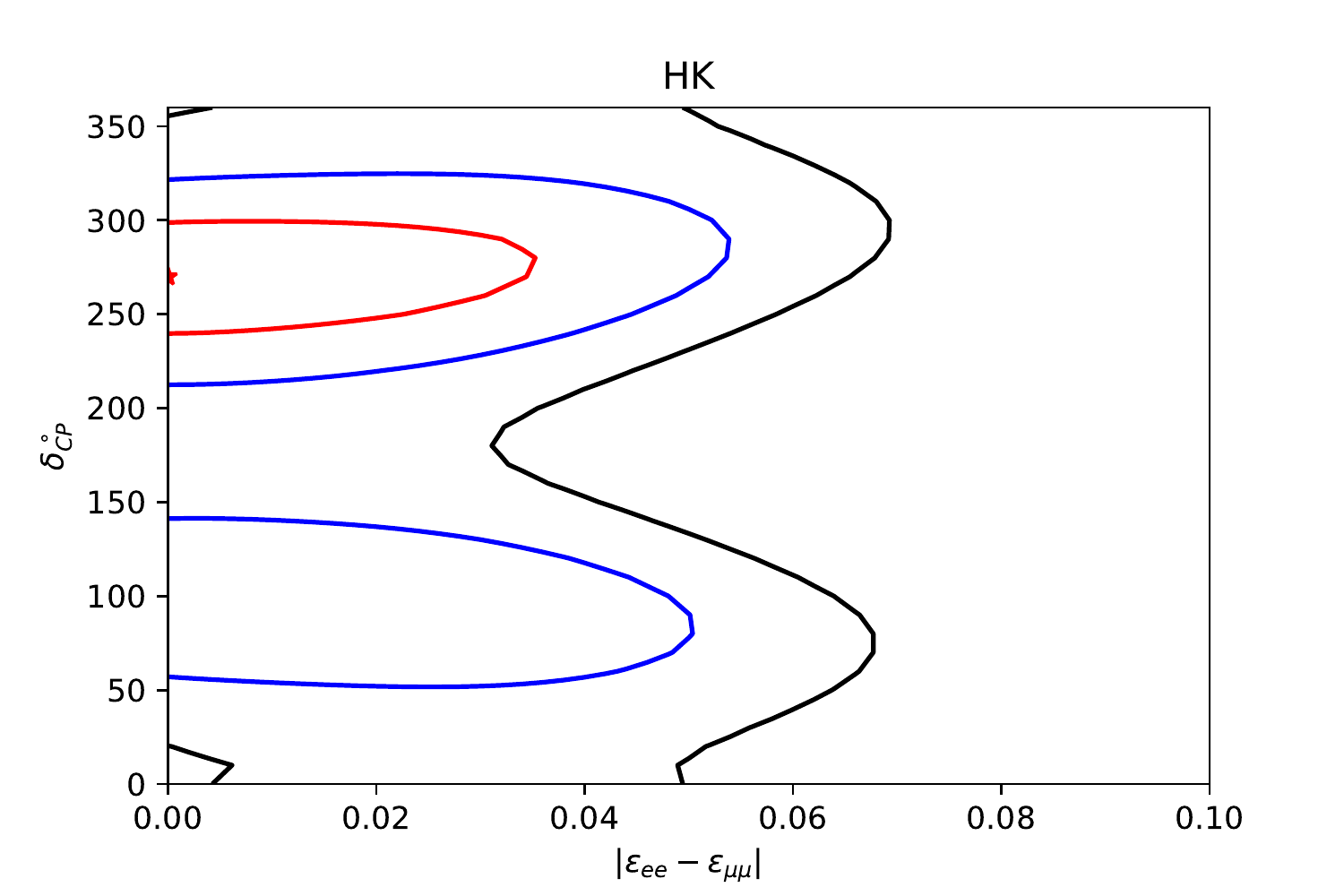}
        \includegraphics[width=0.49\textwidth]{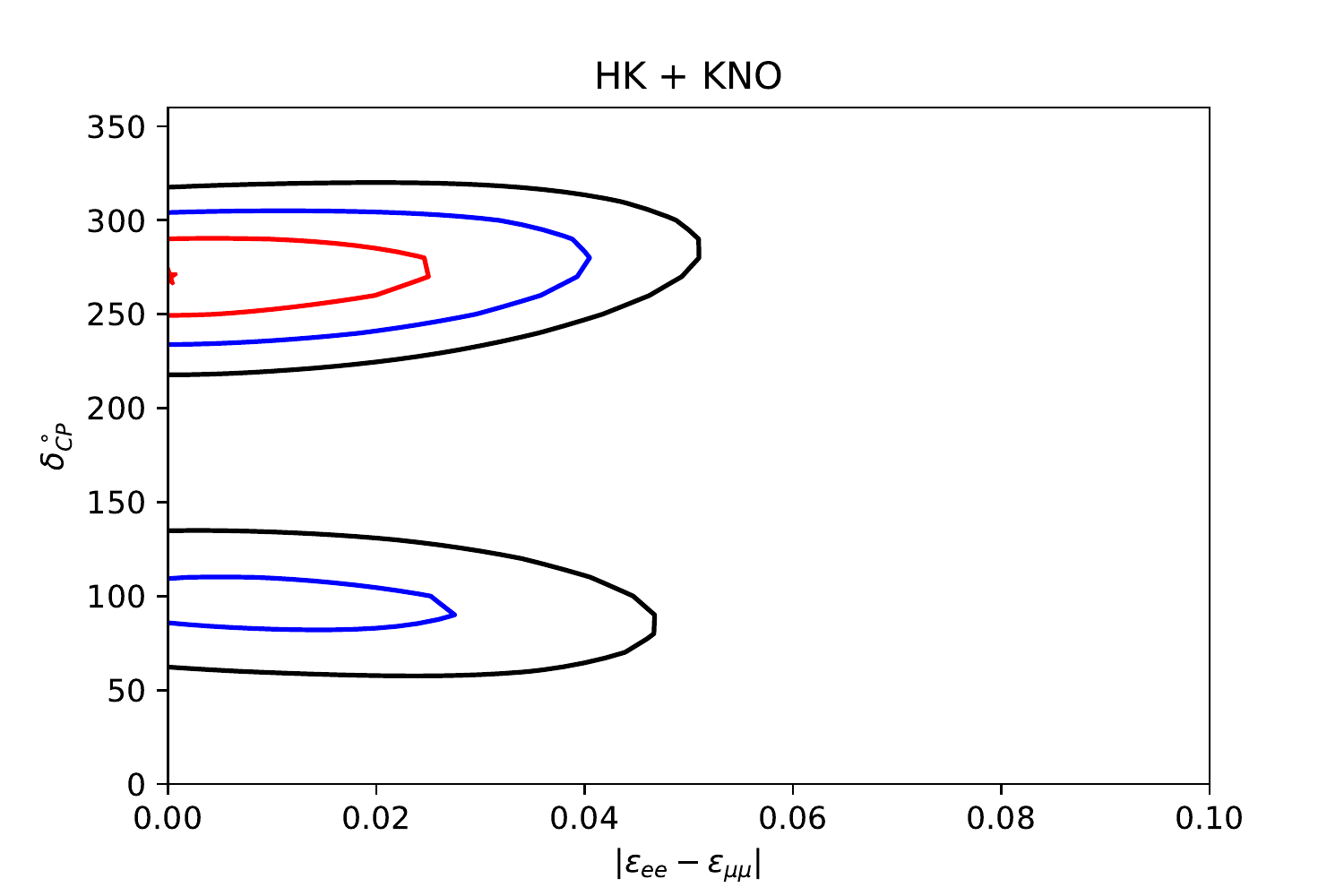}
        \includegraphics[width=0.49\textwidth]{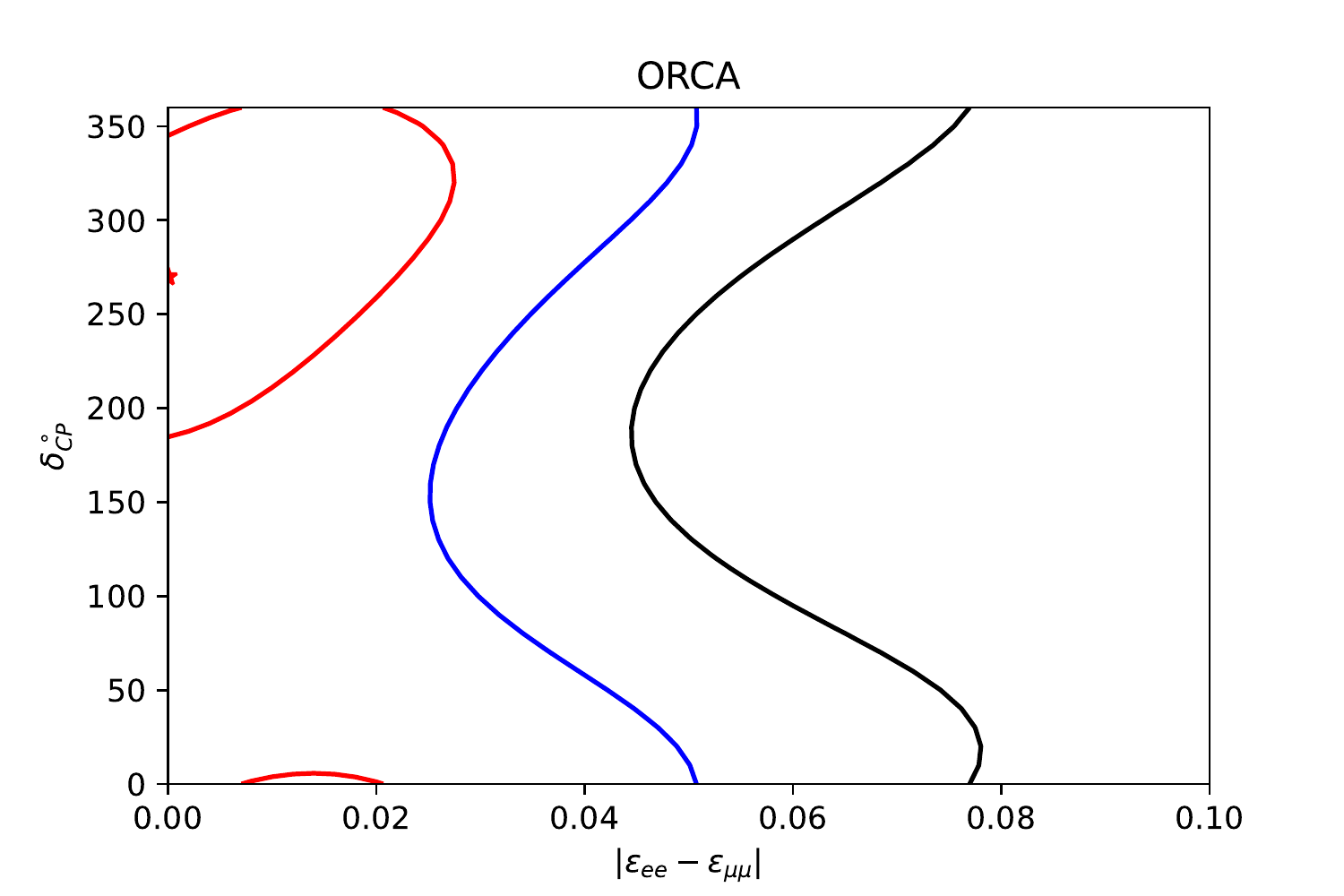} 
 \end{center}
\caption{\label{cord3}   $1\sigma$, $2\sigma$ and $3\sigma$ C.L. contours in the $  \delta_{CP} - | \varepsilon_{e e} -  \varepsilon_{\mu \mu} |$     expected from 10  years of running of DUNE (upper left), HK (upper right), KNO $+$ HK (lower left) and ORCA (lower right)  experiments. We have assumed the true value of $\delta_{CP} = 270^{\circ} $. Known normal neutrino mass hierarchy is assumed to be true. We have assumed the standard model as the true model.     }
\end{figure}

In analyzing the low energy range, we consider seven equal size energy bins: between 1~GeV and 15~GeV.
We have considered electron and muon (anti-)neutrino detection channels. 
Moreover, for $N_{ij} ^{exp}<10$, we assume Poisson distribution function. 
As seen in the figures in the previous section, we investigate the sensitivities of our reference experiments in probing the NSI parameters $\varepsilon_{e\mu}$, $\varepsilon_{e\tau}$, and $| \varepsilon_{ee} - \varepsilon_{\mu \mu} |$ along with their correlations with the CP-violating phase $\delta_{CP}$.

We display the $1\sigma$, $2\sigma$ and $3\sigma$ C.L. sensitivities in the $\delta_{CP} - \varepsilon_{e \mu}$ plane for DUNE, HK, KNO + HK, and ORCA experiments in Fig. \ref{cord}. We have taken the equal size (2 GeV) energy bins between 1 GeV to 15 GeV, except for ORCA where the energy threshold is 3 GeV. We do not consider low energies less than 1 GeV because of the large flux uncertainty and a very small number of events~\cite{Honda:2007uxa}.  

Notice that in this energy range, the cross section of $\nu_{\tau}$ charged-current scattering is smaller than those of $\nu_e$, $\nu_{\mu}$ charged-current scatterings (and neutral-current scattering as well)~\cite{Paschos:2001np,Kretzer:2002fr,Jeong:2010nt,FASER:2019dxq,Messier:1999kj}.
The main source of the background for the $\tau$-neutrino detection is charge miss-identification and neutral current events. According to our calculation, the number of signal events is just $\mathcal O (0.1\%)$ of the expected background events. Thus, the $\tau$-neutrino detection 
in this energy range is quite challenging and 
cannot help to improve the constraint on the parameters.

As we can observe, assuming $\delta_{CP} = 270 ^{\circ}$, KNO combined with the HK data being the most sensitive detector, can exclude
no CP violation at $3\sigma$ C.L.. 
Note that KNO + HK data can determine $\delta_{CP}=270^\circ\pm25^\circ$ within 1$\sigma$ C.L. . 
This is due to its large volume but small $E_{\rm th}$ compared to ORCA.
Note that HK alone and DUNE can nevertheless exclude no CP violation hypothesis at $2\sigma$ C.L. and $1\sigma$ C.L., respectively, while ORCA cannot exclude it.
Comparing with the current global analysis results ~\cite{Esteban:2019lfo,Esteban:2020cvm} which have no sensitivity to $\delta_{CP}$ at $1\sigma$, the future experiments can exclude no CP violation ($\delta_{CP} = 0,~180^\circ$) up to $3\sigma$ C.L..
In addition, we expect the ten years of data taking at DUNE, HK, KNO + HK can improve the sensitivities on $\varepsilon_{e \mu}$, $\varepsilon_{e \tau}$ and $| \varepsilon_{ ee } - \varepsilon_{ \mu \mu }|$ by 2 times, 4 times and one orders of magnitude, respectively, compared to the current bounds~\cite{Esteban:2019lfo}.

In a similar way, the sensitivities on the $\delta_{CP} - \varepsilon_{e \tau}$ plane for DUNE, HK, KNO + HK, and ORCA experiments are presented in Fig.~\ref{cord2}.
The sensitivity of each experiment looks similar to that in Fig.~\ref{cord}.
Note that the experiments are rather sensitive to probing the positive values of $\varepsilon_{e\tau}$ in contrast to the negative values of $\varepsilon_{e\mu}$ by comparing Fig.~\ref{cord} and \ref{cord2}.
This is because $\varepsilon_{e\mu}$ and $\varepsilon_{e\tau}$ give different contribution to $P_{e \mu}$ with different signs~\cite{Kopp:2007ne}.
We also show the allowed regions in  $\delta_{CP} - | \varepsilon_{e e} - \varepsilon_{\mu \mu} |$ plane for each experiment in Fig. \ref{cord3}. 
As seen in the previous section, the correlation is much weaker than $\delta_{CP} - \varepsilon_{e\mu}$ or $\delta_{CP} - \varepsilon_{e\tau}$.
Nevertheless, we observe that KNO + HK, HK, and DUNE can exclude the no CP violation hypothesis at $3\sigma$, $2\sigma$, and $1\sigma$ C.L., respectively.

\section{Conclusions}
\label{sec:conclusions}

In this paper, we have analyzed the potential of future large-volume neutrino experiments in probing the non-standard interactions from atmospheric neutrino data.
As reference experiments fulfilling our purpose, we choose DUNE, HK, KNO (combined with the HK data), and ORCA assuming the ten years of data taking.
For DUNE, we consider the $\nu_\tau$ appearance due to its ability to detect the $\nu_\tau$ charged current scattering signal event by event assuming its detection efficiency is 30\%, while only the $\nu_\mu$ disappearance is considered in the other experiments.
Since the oscillation probabilities of $\nu_\mu \to \nu_\tau$, $\nu_\mu \to \nu_\mu$, $\nu_e \to \nu_e$, $\nu_e \to \nu_\mu$
(and their CP conjugate processes) depend on the oscillation parameters differently with the neutrino energy, we separately analyze the high energy range, i.e., $E \in [15, 200]$ GeV, and the low energy range, i.e., $E \in [1, 15]$ GeV.
For simplicity, we also assume that the mass hierarchy is known as normal ordering.

In the high energy range, all the experiments have the sensitivities probing $\varepsilon_{\mu \tau}$ around $(1.0 - 5.5) \times 10^{-4}$ and $|\varepsilon_{\mu \mu} - \varepsilon_{\tau \tau}|$ around $(0.9 - 3.9) \times 10^{-3}$, which are about two orders of magnitude improvements from the current bounds.
The size of the fiducial volume takes a key role in the analysis so better sensitivities are obtained from ORCA, KNO + HK, HK, and DUNE in order.
Nevertheless, the combination of the $\nu_\tau$ appearance and $\nu_\mu$ disappearance renders DUNE have competitive sensitivities in probing $\varepsilon_{\mu \tau}$ to the other much larger size experiments.
Therefore it is important to develop new methods to increase the $\nu_\tau$ detection efficiencies both in DUNE or other types of experiments.
Since the energy $E > 15$ GeV is well above the $\theta_{13}$ resonance, the oscillation is in the $2\nu$ approximated system, and hence it is hard to probe $\delta_{CP}$ in this range.

In the low energy range, on the other hand, the neutrino energy can be below the $\theta_{13}$ resonance so that the experiments can have sensitivities on $\delta_{CP}$.
We represented our analysis results in the planes of $\delta_{CP} - \varepsilon_{e\mu}$, $\delta_{CP} - \varepsilon_{e\tau}$, and $\delta_{CP} - |\varepsilon_{ee} - \varepsilon_{\mu \mu}|$.
The experimental sensitivities rely on both the size of the fiducial volume and the threshold energy of each experiment in the low energy range.
Therefore KNO combined with the HK data is expected to have the best sensitivities and can determine $\delta_{CP}=270^\circ\pm25^\circ$ within  1$\sigma$ C.L.. 
Note that the data from KNO can have less background due to its 1000 m scale granite overburden and the actual sensitivity can be enhanced from our estimation here, which will be discussed in our future work.
We expect that HK alone and DUNE can exclude no CP violation hypothesis at $2\sigma$ C.L. and $1\sigma$ C.L., respectively, while ORCA cannot exclude it due to its rather high threshold energy (3 GeV).  
However, if the energy threshold of ORCA  reduces to lower energies, its sensitivity to the determination of $\delta_{CP}$ will increase significantly. It is worth mentioning that, since DeepCore has a comparable fiducial volume to ORCA and its energy threshold is 5 GeV (with a future plan of lowering it)~\cite{IceCube:2011ucd}, our nominal 
results can be applicable to DeepCore too. 
Our analysis results show that various future neutrino experiments have intriguing potential in probing the CP violation in correlation with the NSI parameters $\varepsilon_{e\mu}$, $\varepsilon_{e\tau}$, and $|\varepsilon_{ee} - \varepsilon_{\mu \mu}|$: at $3\sigma$ C.L. for $\delta_{CP}$ and $\mathcal O (10^{-2})$ for the NSI parameters, 
which is an order of magnitude improvements from the current bounds.

Consequently, we expect the future improvements of the $\nu_\tau$ detection efficiency in the experiments such as DUNE and lowering $E_{\rm th}$ in gigantic size experiments such as ORCA and DeepCore are crucial in probing the NSI parameters and the CP violation phase.
Further dedicated studies by experimental colleagues are highly encouraged.
Although not implemented here, the analysis assuming the unknown mass ordering and combining with the long-baseline data would provide more interesting results, which we leave as future work.

\subsection*{Acknowledgments}
Authors are grateful to Alexei Yu. Smirnov, Siyeon Kim, Yu Seon Jeong and Chang Hyon Ha for useful remarks. PB and MR are grateful to Chung-Ang University for the warm hospitality.
Hospitality at APCTP during the program “Dark Matter as a Portal to New Physics” is kindly acknowledged.
This work is supported by the National Research Foundation of Korea (NRF-2020R1I1A3072747).

\appendix

\section{The possible enhancement of $\nu_\tau$ detection efficiency}
\label{sec:appen}

\begin{figure}
\begin{center}
\includegraphics[width=0.49\textwidth]{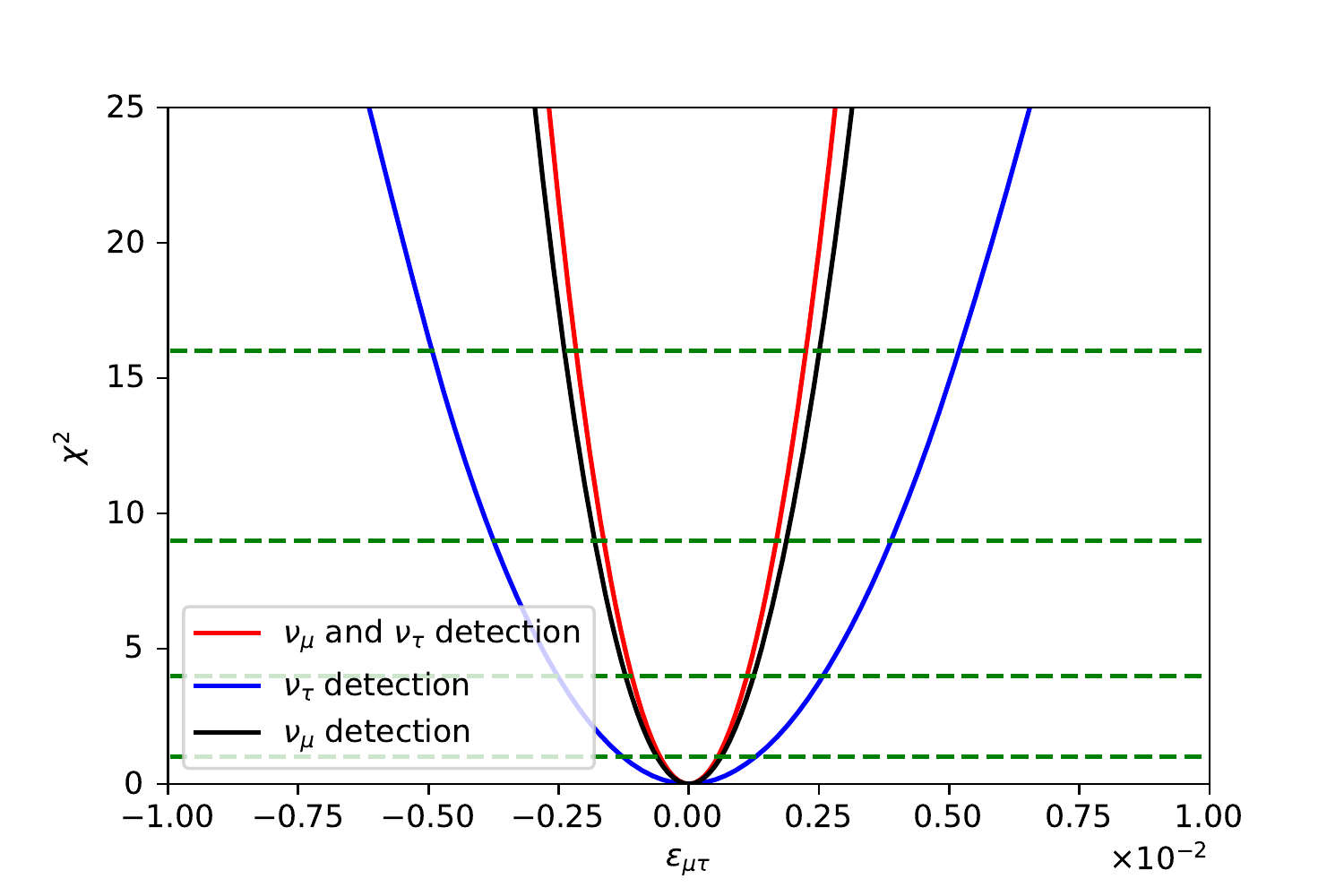}
\includegraphics[width=0.49\textwidth]{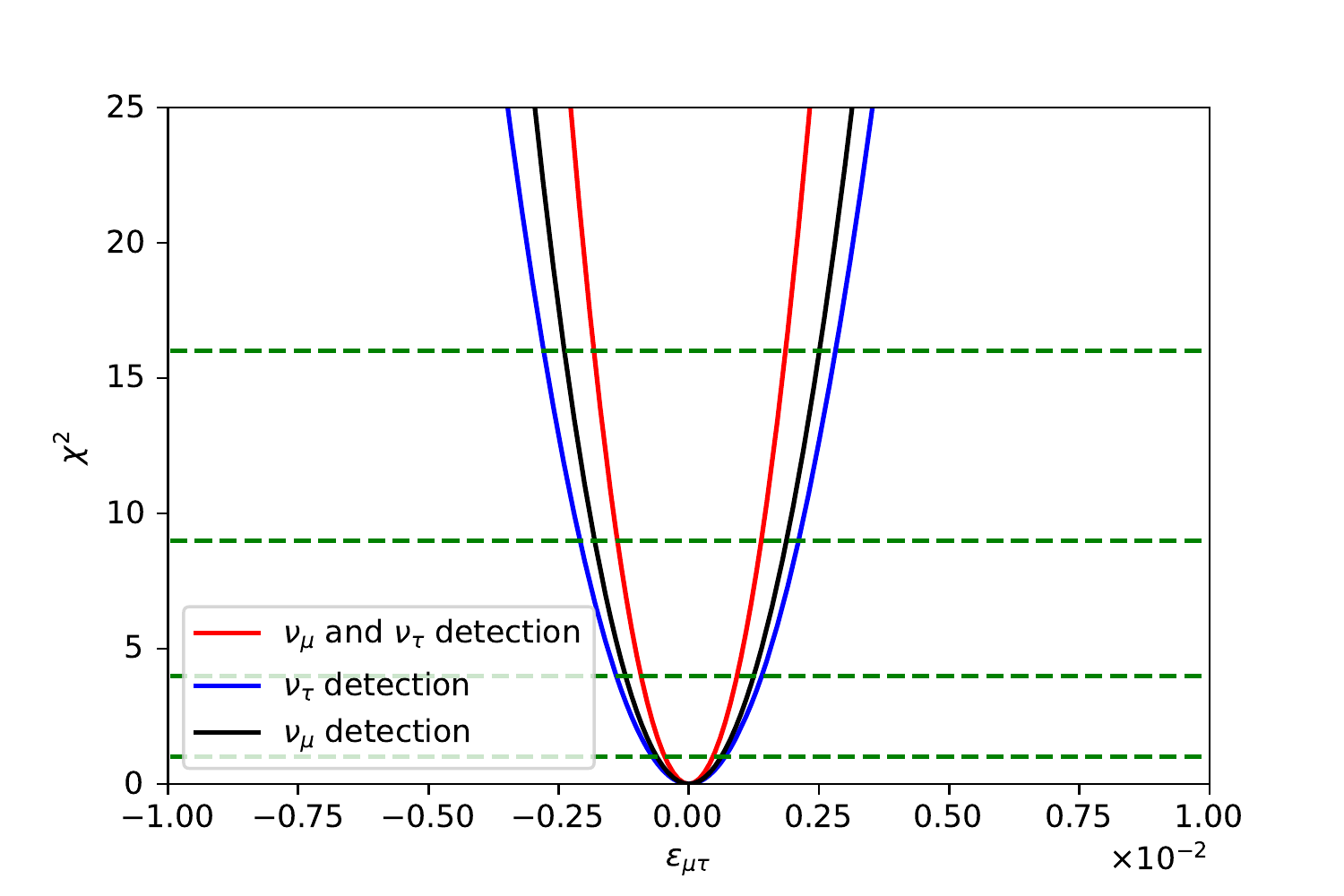}
\includegraphics[width=0.49\textwidth]{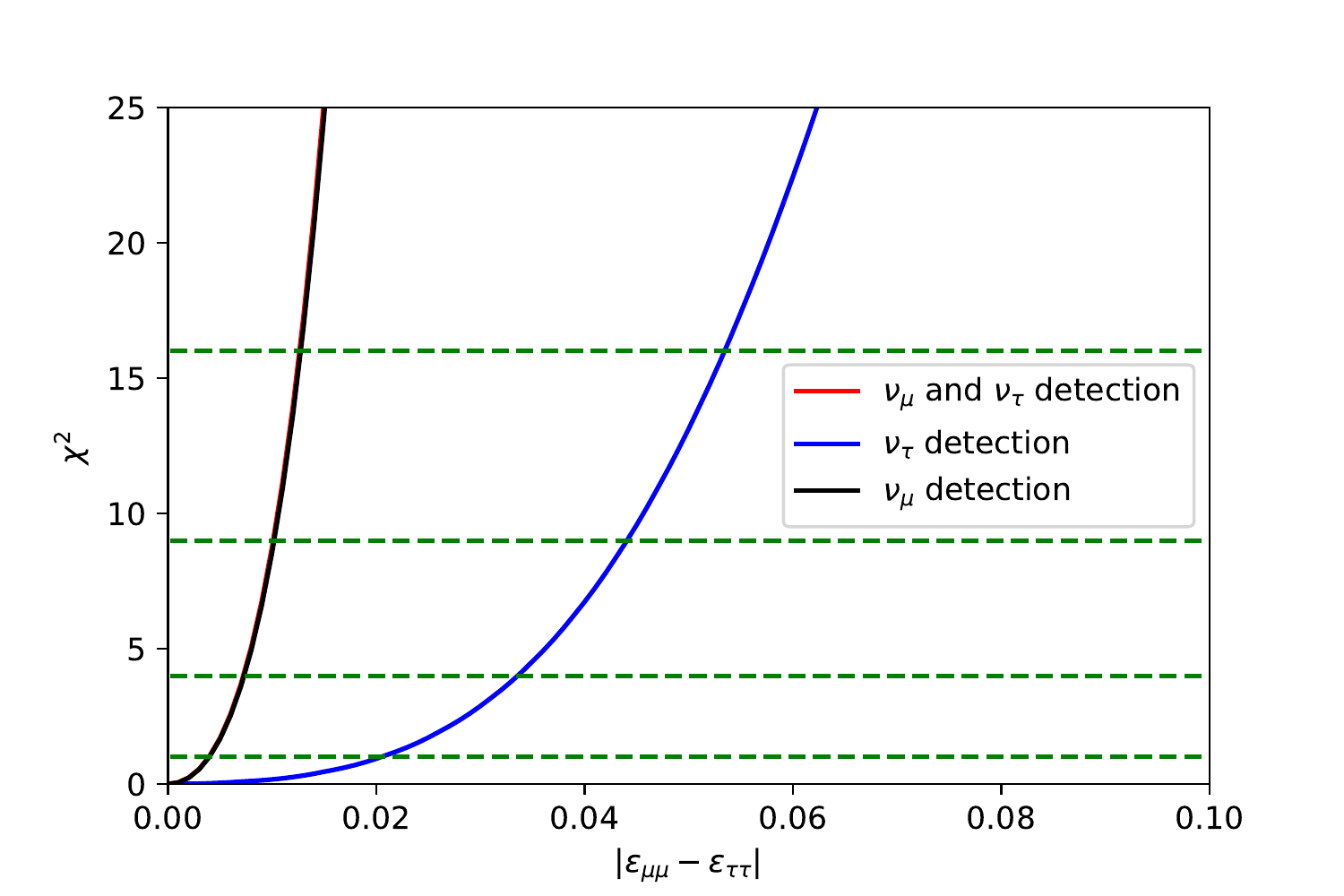}
\includegraphics[width=0.49\textwidth]{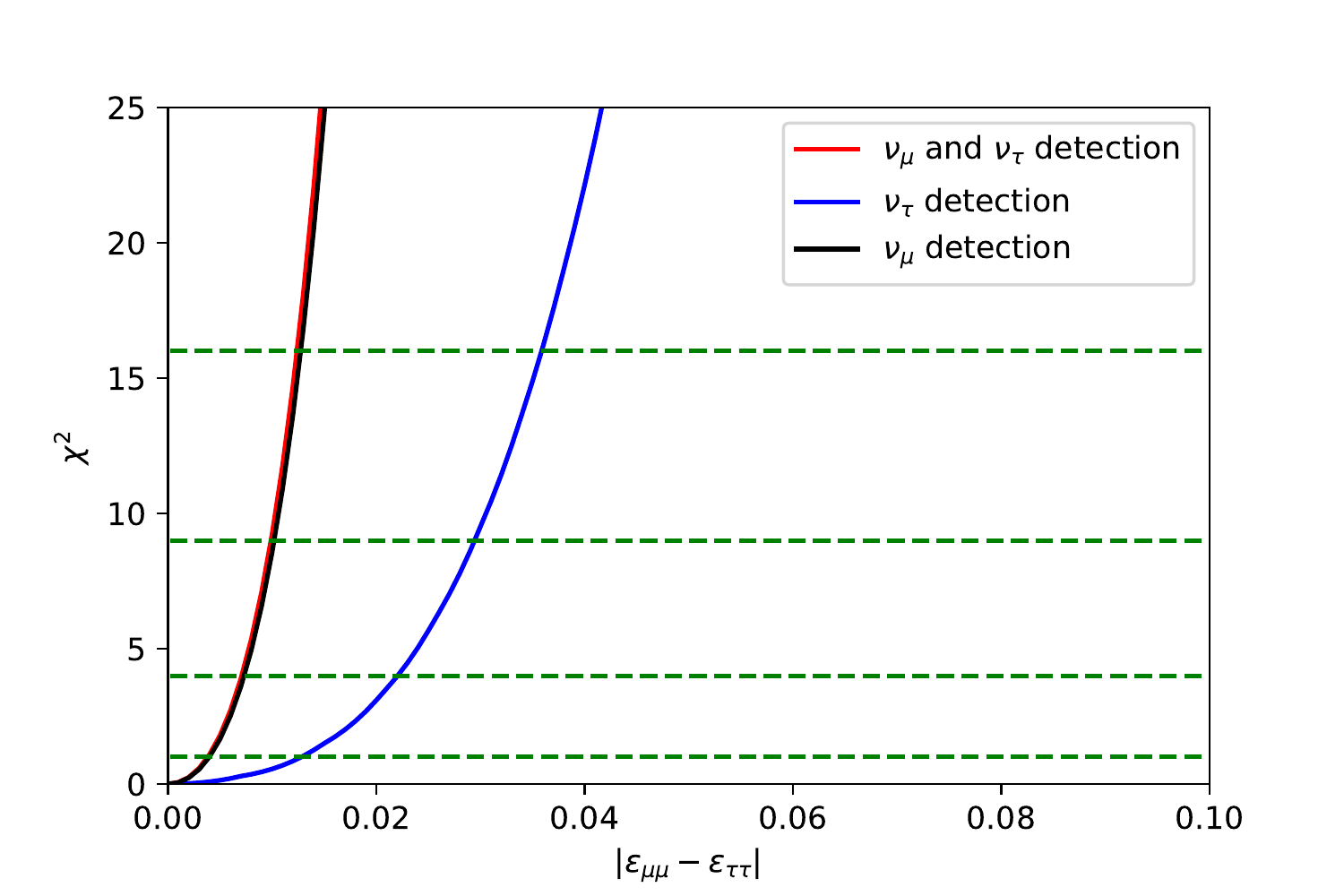}
\end{center}
\caption{\label{chi11} The expected $\chi$-squared
for NSI as a function of $ | \varepsilon_{\mu \tau} | $ and $ | \varepsilon_{\mu \mu} - \varepsilon_{\tau \tau} |$ assuming ten years of data taking by DUNE. The blue and the black curves correspond to $\nu_ \tau$ and $\nu_ \mu$ detection channels, respectively. The red curve corresponds to the sum of $\nu_ \tau$ and $\nu_ \mu$ detection channels.
We have assumed detection efficiency of $\nu_ \tau$ as $30\% $ (left panels) and $100\% $ (right panels).
The neutrino mass hierarchy is assumed to be known and normal. The horizontal dashed green lines show $1\sigma$, $2\sigma$, $3\sigma$ and $4\sigma$ lines.
}
\end{figure}

Since DUNE can detect $\tau$-neutrinos, we show the impact of increasing the detection efficiency in probing the NSI parameters in DUNE-like experiments.
In Fig.~\ref{chi11}, we present $\chi^2$ after the 10 years of running of the DUNE experiment as a function of the NSI parameters $\varepsilon_{\mu\tau}$ (upper plots) and $| \varepsilon_{\mu \mu} - \varepsilon_{\tau \tau} |$ (lower plots). We assume the NSI parameters other than that shown in the horizontal axis in each panel are zero. 

The $\nu_\tau$ detection efficiency is assumed to be $30\%$ in the left panels and $100\%$ in the right panels. 
As mentioned in \cite{DUNE:2020ypp}, for $\tau$ neutrinos the efficiency is equal to $30\%$ while for higher energies we assume a perfect efficiency.

The blue and black curves correspond to the $\nu_{\tau}$ and $\nu_{\mu}$ detections, respectively. 
The red curve corresponds to including $\nu_{\tau}$ and $\nu_{\mu}$ detection together. 
The horizontal dashed green lines correspond to the $1\sigma$, $2\sigma$, $3\sigma$ and $4\sigma$ lines. 
As can be observed from comparing the left and the right panels, the increase of the $\nu_\tau$ detection efficiency from 30\% to 100\% notably enhances the impact of adding the $\nu_\tau$ appearance in probing the $\varepsilon_{\mu \tau}$, although it is not so notable for $|\varepsilon_{\mu \mu} - \varepsilon_{\tau \tau}|$.

\bibliography{ref}

\end{document}